\documentstyle[12pt,a4,epsf]{article}
\newcommand{\rcite}[1]{{\cite{#1}}}
\newcommand{\rref}[1]{{(\ref{#1})}}
\newcommand{\tref}[1]{{\ref{#1}}}
\newcommand{\rlabel}[1]{{\label{#1}}}
\newcommand{\rbibitem}[1]{\bibitem{#1}}
\newcommand{\be}{\begin{equation}}
\newcommand{\ee}{\end{equation}}
\newcommand{\ba}{\begin{eqnarray}}
\newcommand{\ea}{\end{eqnarray}}
\newcommand{\dis}{\displaystyle}
\newcommand{\fourp}{\Pi^{\rho\nu\alpha\beta}(p_1,p_2,p_3)}
\renewcommand{\mathrm}[1]{{\rm #1}}
\newcommand{\tr}{\mathrm{tr}}
\def\theequation{\arabic{section}.\arabic{equation}}

\begin{document}
\begin{titlepage}
\begin{flushright}
{Revised}\\
{NORDITA-95/75 N,P}\\
{hep-ph/9511388}
\end{flushright}
\vspace{2cm}
\begin{center}
{\large\bf Analysis of the Hadronic Light-by-Light\\
 Contributions to the Muon $g-2$}\\
\vfill
{\bf Johan Bijnens$^a$, Elisabetta Pallante$^a$
 and Joaquim Prades$^{a,b}$
\footnote{Present address: Departament de
 F\'{\i}sica Te\`orica, Universitat de Val\`encia and
Institut de F\'{\i}sica Corpuscular, CSIC -
Universitat de Val\`encia,
 C/ del Dr. Moliner 50, E-46100 Burjassot (Val\`encia), Spain.}
}\\[0.5cm]
${}^a$ NORDITA, Blegdamsvej 17,\\
DK-2100 Copenhagen \O, Denmark\\[0.5cm]
$^b$ Niels Bohr Institute, Blegdamsvej 17,\\
DK-2100 Copenhagen \O, Denmark\\[0.5cm]

\end{center}
\vfill
\begin{abstract}
 We calculate the hadronic light-by-light contributions
to the muon $g-2$. We use both
$1/N_c$ and chiral counting to organize the calculation. Then
we calculate the leading and next-to-leading order in the
$1/N_c$ expansion low energy contributions using the Extended
Nambu--Jona-Lasinio model as hadronic model. We do that to all orders
in the external momenta and quark masses expansion.
Although the hadronic light-by-light
contributions to muon $g-2$ are not saturated by these
low energy contributions we estimate them conservatively.
A detailed analysis of the different hadronic light-by-light
contributions to muon $g-2$ is done. The dominant contribution
is the twice anomalous pseudoscalar exchange diagram.
The final result we get is $a_\mu^{\rm light-by-light}=
 (-9.2\pm3.2 )  \cdot 10^{-10}$.
This is between two and three times the expected experimental uncertainty
at the forthcoming BNL muon $g-2$ experiment.
\end{abstract}
\vspace*{1cm}
\vfill
 March 1996
\end{titlepage}

\section{Introduction}
\rlabel{1}
\setcounter{equation}{0}

The forthcoming experiment at Brookhaven National Laboratory (BNL)
 \rcite{RO92} plans to measure the anomalous magnetic moment
of the muon $a_\mu \equiv (g_\mu - 2)/2$ with an accuracy around
$\pm \, 4 \cdot 10^{-10}$, improving by a factor more than
twenty the previous experimental determination at CERN \rcite{PDG94}:
\be
a_\mu^{\rm exp} = 11 \, 659 \, 230 (84) \cdot 10^{-10} .
\ee
This expected impressive performance
has motivated the recent raised interest  in obtaining a more accurate
theoretical prediction of $a_\mu$ within the Standard Model (SM),
for reviews see \rcite{gm2}.
One of the reasons is that with a theoretical uncertainty of the
same order as the aimed  BNL uncertainty, the anomalous
magnetic moment of the muon could become a precision test of
the quantum corrections of the
electroweak sector of the Standard Model.
The one-loop electroweak contributions to $a_\mu$ are \rcite{oneloop}
\be
\rlabel{oneEW}
a_\mu^{\rm EW}(\rm one-loop)= 19.5 (0.1) \cdot 10^{-10} .
\ee
These are order $m^2 G_F$ with $G_F$ the Fermi coupling constant and
$m$ the muon mass.
Electroweak contributions from two-loops \rcite{KKSS92} of
order $m^2 G_F (\alpha/\pi)$
have been recently reanalyzed by \rcite{CKM95,PPR95} for
the fermionic part
and by \rcite{CKM95b} for the bosonic one,  with the result
\rcite{CKM95b}
\be
\rlabel{twoEW}
a_\mu^{\rm EW}({\rm two-loops})= -4.4 (0.4) \cdot 10^{-10} .
\ee
With the same low theoretical and experimental uncertainties, and
when combined with other high precision results,
{}from the LEP experiments and elsewhere,
$a_\mu$ would become an excellent probe of physics beyond the Standard Model
(extra W$_{\rm R,L}$ gauge bosons,
Z' bosons, extra Higgs bosons, SUSY, $\ldots$). For a study
of the sensitivity of $a_\mu$ to new physics see  for instance
\rcite{KM90}.

In the Standard Model, the contributions to $a_\mu$
fall into three categories: the pure electromagnetic (QED)
contributions, the electroweak contributions discussed
above and the hadronic
contributions. The QED contributions have been calculated and/or
estimated up to order $(\alpha/ \pi)^5$.
They give the bulk of the value of $a_\mu$.
For an updated value of $a_\mu^{\rm QED}$ see \rcite{CKM95b}
and references therein. A review of the QED calculations is in
\rcite{KM90}. The result is
\be
\rlabel{QED}
a_\mu^{\rm QED} = 11 \, 658 \, 470.6 (0.2) \cdot 10^{-10} \, .
\ee
The main actual source of theoretical uncertainty still remains in the
hadronic contributions. The leading hadronic contributions
are of two types: the vacuum polarization and the
light-by-light scattering contributions.
The hadronic vacuum polarization contributions are the
major source of uncertainty at present. Fortunately,
they can be related through a dispersion
relation to the experimental ratio $R (s) \equiv \sigma
(e^+e^-\to {\rm hadrons}) /  \sigma (e^+e^-\to \mu^+\mu^-)$
\rcite{BM61}. The planned improvement of the experimental
determination of $R(s)$ in e.g. BEPC at Beijing,
DA$\Phi$NE at Frascati, and VEPP-2M at Novosibirsk
will significantly  reduce this uncertainty.  A
recent reanalysis of the contribution
of the full photon vacuum polarization insertion
into the electromagnetic vertex of the
muon can be found in \rcite{EJ95,AY95}. In \rcite{EJ95},
using experimental data, they find
\be
\label{EJ}
\rlabel{vacpol1}
a_\mu^{\rm vac. pol.} =
725.04 (15.76) \cdot 10^{-10} .
\ee
In \rcite{AY95}, using experimental data below
2 GeV$^2$ and accurate QCD calculations above this scale, they
find
\be
\label{AY}
\rlabel{vacpol2}
a_\mu^{\rm vac. pol.} =
711.34 (10.25) \cdot 10^{-10} .
\ee
An alternative attempt to compute it at lowest order
$(\alpha / \pi)^2$
within the same low energy model we use in this work can be
found in \rcite{deR94,PA95}. At order $(\alpha / \pi )^3$
there appear other hadronic vacuum polarization contributions
that also can be  expressed as a
convolution of $R(s)$ \rcite{CNPR76}.
They have been estimated to be \rcite{CNPR76,KNO85}
\footnote{This number only contains the $\alpha^3$
corrections not included in \rref{vacpol1} or \rref{vacpol2}.}
\be
\rlabel{vacpolhigh}
a_\mu^{\rm vac. pol.} (\rm higher \, \, orders)
= -19.9(0.4) \cdot 10^{-10} .
\ee

Unfortunately, the hadronic light-by-light scattering  contribution
cannot be related to any observable and
hence we must rely on a purely theoretical framework
to calculate it. There have been several attempts to calculate
this contribution in the past \rcite{CNPR76,KNO85}.
There has also been recently some discussion about the reliability
of this calculation \rcite{BR92,EI94}. To pin down this
contribution is an important issue since a quick estimate
yields that it could be of the same order of magnitude as the expected
BNL uncertainty. Recently, with the aim of reducing as much as possible
the theoretical uncertainty from this contribution,
there have appeared two works, \rcite{HKS95} and
\rcite{BPP95}, which calculate $a_\mu^{\rm light-by-light}$.
An extended version of \rcite{HKS95} is in
\rcite{HKS95b}. This paper is the detailed version of \rcite{BPP95}.

The present work is devoted to
the calculation of  the contributions
of the hadronic light-by-light scattering to $a_\mu$.
A first simplified version and summary of the
main results of this paper was presented in \rcite{BPP95}.
A numerical mistake was discovered in the first reference of
\rcite{BPP95} which was corrected in the Erratum.
Of course, the methods used in \rcite{HKS95} and \rcite{BPP95}
are similar and mainly based
on the analysis performed in \rcite{deR94}.
Nevertheless, we want
to discuss the main differences in the calculations as well as
the reasons why we use the Extended Nambu--Jona-Lasinio (ENJL)
model for this task.

The framework we have adopted to calculate the
hadronic light-by-light contribution is an
$1/N_c$ expansion within the ENJL model. In Ref. \rcite{BPP95}
the ${\cal O} (N_c)$ leading hadronic contributions were presented.
The next-to-leading in the $1/N_c$ expansion effects of the
U(1)$_A$ anomaly were also included. There, we took as a first
estimate the result and error for the ${\cal O}(1)$ in the
$1/N_c$ expansion $\pi^+$ and $K^+$ loop contributions from \rcite{HKS95}.
Here we shall come back to all these issues in a more detailed
fashion.

The paper is organized as follows. In Section \tref{2} we present
the definitions related with the hadronic
light-by-light scattering contribution
to $a_\mu$ and specify the method we have used to
calculate it. In Section \tref{3} we
explain why we have chosen the ENJL model
as a good low energy hadronic model. Its main
features and definitions needed are also presented
here. Section \tref{4} concerns the calculation of the
large $N_c$ contributions to the hadronic light-by-light
scattering to $a_\mu$. This section also includes an estimation
of the main next-to-leading in $1/N_c$ effects
coming from the U(1)$_A$ anomaly. Various checks performed
and numerical comparison with other works are also
shown here.
In Section \tref{5} the next-to-leading (${\cal O} (1)$)
in $1/N_c$ contributions coming from charged pion and
kaon loops are discussed. Here,
inspired by the ENJL model ${\cal O}(N_c)$ calculation,
we will use lowest order Chiral Perturbation
Theory (CHPT) modulated  with vector meson propagators
to calculate them. We discuss issues of gauge and chiral
 invariance there as well.
Then in Section \tref{6} we shall gather the numerical results
for the contributions analyzed in the previous sections.
In Section \tref{7} we discuss
the contributions coming from the intermediate (between 0.5 GeV and 4 GeV)
and higher energy regions
to the hadronic light-by-light  scattering contribution to $a_\mu$.
In particular, how to estimate them and the theoretical error they induce.
Finally, Section \tref{8} is devoted to a discussion
of the results and conclusions. Appendices collecting some analytical
expressions are also included at the end.

\section{The Method and Definitions}
\rlabel{2}
\setcounter{equation}{0}

The amplitude describing the interaction of a momentum $p$
fermion with an external electromagnetic field $A^\mu$
with momentum transferred $p_3\equiv p - p'$, can be written as
\be
\rlabel{f1f2}
{\cal{M}} \equiv
-\vert e\vert A^\mu \bar{u}(p^\prime )\left[
F_1(p_3^2)\gamma_\mu
-F_2(p_3^2) i{\sigma_{\mu\nu}\over 2m}p_3^\nu
-F_3(p^2_3)
\gamma_5 {\sigma_{\mu\nu} \over 2m}p_3^\nu \right] u(p),
\ee
where $m$ is the fermion mass.
The form factor $d\equiv-F_3(0)$ is the electric dipole moment
and $\mu \equiv F_1(0)+F_2(0)$ is the magnetic moment of the
fermion in  magnetons.
In the Born approximation
$F_1(0)=1$ and $F_2(0)=F_3(0)=0$. In analogy with
the classical limit, it is usual to  define the
gyromagnetic ratio $g\equiv 2 \mu$ and
the anomalous magnetic moment as $a \equiv (g -2)/2$.

\begin{figure}
\begin{center}
\leavevmode\epsfxsize=10cm\epsfysize=8cm\epsfbox{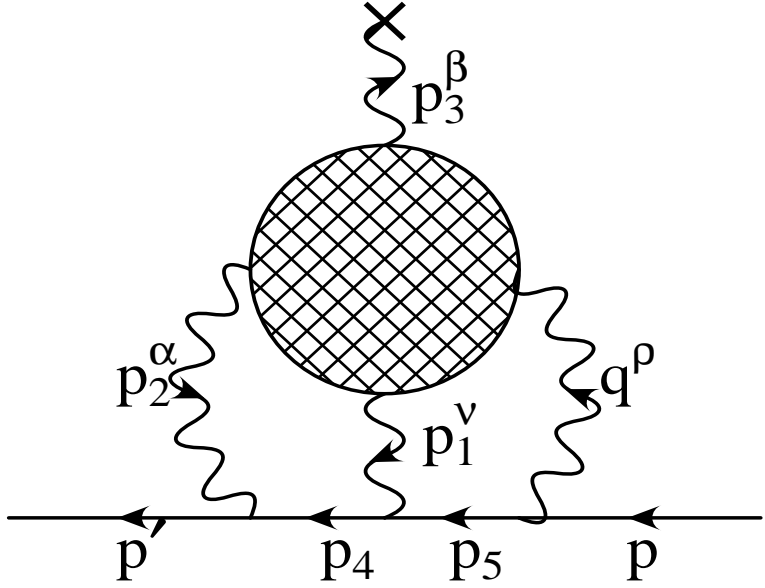}
\end{center}
\caption{ Hadronic light-by-light contribution to $a_\mu$.
The bottom line is the
muon line. The wavy lines are photons and the cross-hatched circle depicts
the hadronic part. The cross is an external vector source.}
\label{fig1}
\end{figure}

The hadronic light-by-light scattering contributes to $a$
at order $(\alpha / \pi)^3$.  This is a vector
four-point function made out of four quark vector currents
attached to the fermion line with three of its legs coupling to photons
in all possible ways and the fourth vector leg coupled to the
electromagnetic external source.  One hadronic
light-by-light scattering contribution is depicted
in Figure \tref{fig1}.
The cross-hatched blob is the hadronic part.
The momenta in this figure correspond to the first permutation
of the three vector legs attaching to the fermion.
There are five more permutations.

To extract the hadronic light-by-light scattering contribution to
the muon $a_\mu \equiv (g_\mu -2)/2$
we shall closely follow Ref. \rcite{ABDK70}
and use the first permutation shown in Fig. \tref{fig1} as an example.
The hadronic light-by-light  contribution to ${\cal M}$ is
\ba
\rlabel{amplitude}
{\cal M}&=&\vert e\vert^7 A_\beta
\int {{\rm d}^4 p_1 \over (2\pi )^4} \int
{{\rm d}^4p_2\over (2\pi )^4} \,
{1\over q^2\, p_1^2 \, p_2^2 (p_4^2-m^2) \,
(p_5^2 - m^2)}\nonumber\\
&\times&  \Pi^{\rho\nu\alpha\beta} (p_1,p_2,p_3) \,
\bar{u}(p^\prime )\gamma_\alpha (\not{\! p}_4 +m )\gamma_\nu
(\not{\! p}_5 +m ) \gamma_\rho u(p) \nonumber \\
&+& {\rm five \,\, more \, \, permutations}\, , \nonumber \\
\ea
where we have explicitly given the contribution from
the permutation in Fig. \tref{fig1}.
In Eq. \rref{amplitude}, $m$ is the mass of the muon and
$\Pi^{\rho\nu\alpha\beta} (p_1,p_2,p_3)$ is the four-point function
\ba
\rlabel{fourpoint}
\Pi^{\rho\nu\alpha\beta} (p_1,p_2,p_3) \equiv
{\dis \sum_{a,b,c,d}} \Pi^{\rho\nu\alpha\beta}_{abcd}
(p_1,p_2,p_3) \equiv \hspace*{4cm}\hfill \nonumber \\
{1 \over 6}\,
{\dis \sum_{a,b,c,d}} \, i^3 \int {\rm d}^4 x \int {\rm d}^4 y
\int {\rm d}^4 z \,
e^{i(p_1 \cdot x+p_2 \cdot y + p_3 \cdot z)} \,
\langle 0 | T \left( V_a^\rho(0) V_b^\nu(x) V_c^\alpha(y)
V_d^\beta(z) \right) | 0 \rangle . \nonumber \\
\ea
$V_i^\mu(x) \equiv Q_i \, \left[
\bar q_i(x) \gamma^\mu q_i (x) \right]$,
$q_i$ is a quark of flavour $i$ and $Q_i$ its
electric charge in units of $|e|$.
Summation over colour between brackets
is understood. The explicit calculation of
$\Pi^{\rho\nu\alpha\beta} (p_1,p_2,p_3)$ in the Standard Model is the
subject of the next sections.

The reason for the factor
$1/6$ in the definition of $\fourp$ is the
following. This four-point function has six contributions
due to Bose symmetry of the vector legs, but these
permutations are just giving the same six permutations we
already took into account in Eq. \rref{amplitude}.
What we do is to calculate with both
the six permutations of the four-point function vertices
and the six permutations
of the points where photons connect to the muon line
and therefore we have to divide by six the four-point function
not to make double counting. This will be very useful since
we can now use both U(1) gauge invariance in the
construction of $\fourp$ (see App. \tref{A})
 and in the electromagnetic vertices on the muon line.

Because of the U(1) gauge invariance,
the sum over all six permutations contributing to
$\fourp$ has to be UV finite in renormalizable theories as the
Standard Model.
This is  true for the sum of all the permutations of a given
class of contributions to $\Pi^{\rho
\nu\alpha\beta}(p_1,p_2,p_3)$. However, each single
permutation  can be divergent. This is the case for
the one fermion loop contribution where some
permutations are UV logarithmically divergent.
In order not to rely on numerical dangerous cancellations,
we write $\Pi^{\rho\nu\beta\alpha}
(p_1,p_2,p_3)$  in a form
where each permutation is explicitly UV  convergent.
Following Ref. \rcite{ABDK70}, we use the U(1) gauge covariance
condition $p_{3\beta}\Pi^{\rho\nu\alpha\beta} (p_1,p_2,p_3)=0$ to obtain
\be
\rlabel{eq10}
\Pi^{\rho\nu\alpha\lambda}(p_1,p_2,p_3)=
-p_{3\beta} {\delta \Pi^{\rho\nu\alpha\beta}(p_1,p_2,p_3)
\over \delta  p_{3\lambda}} \, .
\ee
The presence of the extra derivative with respect
to $p_3$ makes the rhs of \rref{eq10} explicitly UV finite.
Then $\cal{M}$ can be rewritten as
\be
\rlabel{defm}
{\cal{M}} \equiv - \vert e\vert A_\lambda  \, p_{3\beta} \,
\bar{u} \, (p^\prime )\, M^{\lambda\beta}(p_3) \, u(p),
\ee
where, for the permutation in Eq. \rref{amplitude},
\ba
\rlabel{MLB}
M^{\lambda \beta}(p_3)&=& \vert e\vert^6
\int {{\rm d}^4 p_1 \over (2\pi )^4} \int
{{\rm d}^4p_2\over (2\pi )^4} \,
\, {1\over q^2\, p_1^2 \, p_2^2 (p_4^2-m^2) \,
(p_5^2 - m^2)}\nonumber\\
&\times&
\left[ {\delta \Pi^{\rho\nu\alpha\beta} (p_1,p_2,p_3)  \over
\delta p_{3\lambda}} \right]
\gamma_\alpha (\not{\! p}_4 +m )\gamma_\nu
(\not{\! p}_5 +m ) \gamma_\rho  \, .
\ea
It can be shown, see \rcite{ABDK70}, that the contribution
{}from the light-by-light scattering to $a_\mu$ can be
written as
\be
\rlabel{Damu}
a_\mu^{\rm light-by-light} =
{1\over 48 m} \tr [(\not{\! p} +m )M^{\lambda\beta}(0) \,
(\not{\! p} +m ) [\gamma_\lambda,\gamma_\beta ] ] \, .
\ee
The eight dimensional integral in Eq. \rref{MLB}
of the two loops on muon momenta can be reduced to a five dimensional
one, two moduli and three angles, using the symmetries of the system.
The integration over these variables has been done
in Euclidean space. The  integral in Eq.
\rref{fourpoint} brings in other integration parameters, the latter
integral we also perform in Euclidean space.

The momenta flowing through the three photon legs attached
to the muon line run from zero up to infinity,  covering both
the perturbative and non perturbative regimes of QCD.
These two different regimes are naturally separated by
the scale of the spontaneous symmetry breaking
$\Lambda_\chi\simeq 1$ GeV. In the region between $\Lambda_\chi$
and say (4$\sim$5) GeV, the strong interaction
contributions have to match the perturbative QCD
predictions in terms of quarks and gluons.
Hence, rigorously, one should calculate the low energy
contribution to  $\Pi^{\rho\nu\alpha\beta} (p_1,p_2,p_3)$
and match this result with a perturbative QCD calculation of the
high energy contribution. Here this is technically
rather involved  because of
configurations with both high and low
energy photon momenta. If the scale determining the bulk of the
contributions to $a_\mu^{\rm light-by-light}$
were around the muon mass we could attempt to make a pure
low energy calculation that would saturate it.
Were the contributions from the high energy perturbative region not
negligible we would need a more sophisticated model
suitable both in the low and intermediate energy regions.
We have investigated this issue by putting relevant Euclidean
ultraviolet cut-offs, labelled $\mu$ from now on, on the
moduli of the momenta attaching to the muon line in Eq.
\rref{MLB}.

Since, as said before, the momenta in the photon legs
can run up to infinity any accurate calculation should
incorporate the full external momenta dependence of $\fourp$.
This is clearly beyond the
reach of the present state of the art of CHPT.
Alternatively,
one can rely on a good low energy hadronic model.
We have chosen the  Extended Nambu--Jona-Lasinio
model. Reasons for that choice, definitions and main features
of the model are in the next section.
By using this model, we calculate the low energy
contributions to $a_\mu^{\rm light-by-light}$. We then
study the saturation of $a_\mu^{\rm light-by-light}$ by the physics
at scales below or around $\Lambda$, where $\Lambda$ in our case
is the physical cut-off of the ENJL model.
 Although it will turn out that the contributions
{}from intermediate and high energy regions are not negligible,
we shall be able to give a conservative estimate for them.
This will be explained in Section \tref{7}.

\section {The ENJL Model}
\rlabel{3}
\setcounter{equation}{0}

For recent comprehensive reviews on the NJL \rcite{NJL61}
and the ENJL models \rcite{ENJL75}, see Refs. \rcite{Reviews,Physrep}.
Here, we will only summarize the main features,
notation and reasons why we have chosen this model.
More details and some motivations on the version of ENJL we are
using can be found in \rcite{BBR93,BP94a,BRZ94}.

The good features of the ENJL
model  and its suitability for an accurate
calculation of $a_\mu^{\rm light-by-light}$
were realized in Ref. \rcite{deR94}.
In fact, in this reference the hadronic vacuum polarization
to lowest order $(\alpha/ \pi)^2$ was calculated  within the
same ENJL model we use here, obtaining a
good agreement ($\sim $ 15\%) with the phenomenological result
in Ref. \rcite{EI94}. This accuracy, enough for our purposes,
is the maximum we can expect from our calculation.
One of the conclusions of reference \rcite{deR94}
was that the hadronic vacuum polarization to $a_\mu$ saturates
for energies around 1.5 GeV, which is still
a reasonable scale where the model could be applied
without introducing too much uncertainty.
We shall see that in the light-by-light case the contributions
{}from higher energies are not negligible. More comments
on this are in Section \tref{6}.

\subsection{The Model and Determination of Parameters}
\rlabel{3.1}

The QCD Lagrangian is given by
\ba
\rlabel{QCD}
{\cal L}_{\rm QCD} &=& {\cal L}^0_{\rm QCD} -\frac{1}{4}G_{\mu\nu}
G^{\mu\nu} \, , \nonumber\\
{\cal L}^0_{\rm QCD} &=& \overline{q} \left\{i\gamma^\mu
\left(\partial_\mu -i v_\mu -i a_\mu \gamma_5 - i G_\mu \right) -
\left({\cal M} + s - i p \gamma_5 \right) \right\} q \, .
\ea
Here summation over colour degrees of freedom
is understood and
we have used the following short-hand notation:
$\overline{q}\equiv\left( \overline{u},\overline{d},
\overline{s}\right)$; $G_\mu$ is the gluon field in the
fundamental SU(N$_c$) (N$_c$=number
of colours) representation;
$G_{\mu\nu}$ is the gluon field strength tensor in
the adjoint SU(N$_c$) representation; $v_\mu$, $a_\mu$, $s$ and
$p$ are external vector, axial-vector, scalar and pseudoscalar
field matrix sources; ${\cal M}$ is the quark-mass matrix.
The ENJL model we are using corresponds to
the following Lagrangian
\ba
\rlabel{ENJL1}
{\cal L}_{\rm ENJL} &=& {\cal L}_{\rm QCD}^{\Lambda}
+  2 \, g_S \, {\dis \sum_{i,j}} \left(\overline{q}^i_R
q^j_L\right) \left(\overline{q}^j_L q^i_R\right)
\nonumber\\&&
- g_V \, {\dis \sum_{i,j}} \left[
\left(\overline{q}^i_L \gamma^\mu q^j_L\right)
\left(\overline{q}^j_L \gamma_\mu q^i_L\right) + \left( L \rightarrow
R \right) \right] \,.
\ea
Here $i,j$ are flavour indices, $\Psi_{R,L} \equiv
(1/2) \left(1 \pm \gamma_5\right) \Psi$ and
\be
g_V \equiv  {8 \pi^2 G_V(\Lambda)\over N_c \Lambda^2}
\qquad\mbox{,}\qquad g_S
\equiv   {4 \pi^2 G_S(\Lambda)\over N_c \Lambda^2}\, .
\ee
The couplings $G_S(\Lambda )$ and $G_V(\Lambda )$ are
dimensionless and ${\cal O}(1)$ in the $1/N_c$ expansion and summation
over colours between brackets in \rref{ENJL1} is understood.

The Lagrangian ${\cal L}^{\Lambda}_{\rm QCD}$ includes only
low frequency (less than $\Lambda$)
modes of quark and gluon fields.
The low frequency modes of the  gluon fields can be assumed to be fully
absorbed in the coefficients of the local operators
or alternatively described
by vacuum expectation values of gluonic operators.
So at this level we have two different pictures of this model. One is where
we have integrated out all the gluon degrees of freedom and then
expanded the resulting effective action
in a set of {\bf local} operators with quark fields
keeping only the first non-trivial terms in the expansion.
The other picture is obtained when we only integrate out the short distance
part of gluons and quarks. We then again expand the resulting effective
action in terms of low energy  local
operators with gluon and quark fields.
This is described in \rcite{Physrep,BBR93} and the best fits
there correspond to the first alternative. Therefore,
in the present work we will use \rref{ENJL1}
with all gluonic degrees of freedom integrated out.

This model has three parameters plus the current light quark masses.
The  first three parameters are $G_S$, $G_V$ and the physical
cut-off $\Lambda$ of the regularization that we chose to be
proper-time. Although this regulator breaks in general the
Ward identities we impose them by adding the necessary counterterms
(both in the anomalous \rcite{BP94b} and in the non anomalous
sectors). The light quark masses in ${\cal M}$ are fixed in order to
obtain the physical pion and kaon masses in the poles of the
pseudoscalar two-point functions \rcite{BP94a}.
The values of the other parameters are fixed from the results of
the fit to low energy effective chiral Lagrangians obtained
in \rcite{BBR93}. They are $G_S \simeq 1.216$, $G_V
\simeq 1.263$, and $\Lambda=1.16$ GeV from Fit 1 in that reference.
 Solving the gap equation, we then obtain the
constituent quark masses: $M_u=M_d=275$ MeV and $M_s=427$ MeV.

\subsection{Advantages and Disadvantages of the Model}
\rlabel{3.2}

The ENJL model is a very economical model that captures
in a simple fashion
a lot of the observed low and intermediate energy phenomenology. It
has also a few theoretical advantages.

\begin{enumerate}
\item The model in Eq. \rref{ENJL1} has the same
symmetry structure as the QCD action
at leading order in $1/N_c$ \rcite{tH74}. Notice that the
U(1)$_A$ problem is absent at this order \rcite{WI79}.
(For explicit symmetry properties under SU(3)$_L$ $\times$
SU(3)$_R$ of the fields in this model
see reference \rcite{BBR93}.)  In the chiral limit and for $G_S>1$
this model breaks chiral symmetry spontaneously
via the expectation value of the scalar quark-antiquark one-point function
(quark condensate).
\item It has very few free parameters. These are unambiguously
determined from low energy physics involving only pseudo-Goldstone
bosons degrees of freedom\rcite{BBR93}.
\item It only contains constituent quarks. Therefore, all the
contributions to a given process (in particular to $a_\mu$)
are uniquely distinguished using only constituent quark diagrams.
Within this model there is thus no possible double counting.
In particular, the constituent quark-loop contribution and what
would be the equivalent of the meson loop contributions in this model,
are of different order in the $1/N_c$ counting \rcite{deR94}.
This shows clearly that there is no double counting here.
As described in \rcite{BBR93} this model includes
the constituent-quark loop model as a specific limit.
\item Two-point functions are given by the general graph
depicted in Fig.  \tref{fig2pt}.
\begin{figure}
\begin{center}
%
%
%
\thicklines
\setlength{\unitlength}{1mm}
\begin{picture}(140.00,35.00)(0.,15.)
\put(97.50,35.00){\oval(15.00,10.00)}
\put(103.00,33.50){$\bigotimes$}
\put(17.50,35.00){\oval(15.00,10.00)}
\put(25.00,35.00){\circle*{2.00}}
\put(32.50,35.00){\oval(15.00,10.00)}
\put(40.00,35.00){\circle*{2.00}}
\put(47.50,35.00){\oval(15.00,10.00)}
\put(55.00,35.00){\circle*{2.00}}
\put(62.50,35.00){\oval(15.00,10.00)}
\put(08.00,33.50){$\bigotimes$}
\put(68.00,33.50){$\bigotimes$}
\put(88.00,33.50){$\bigotimes$}
\put(38.50,19.00){(a)}
\put(95.50,19.00){(b)}
\put(14.50,40.00){\vector(1,0){3.00}}
\put(29.50,40.00){\vector(1,0){3.50}}
\put(44.00,40.00){\vector(1,0){5.00}}
\put(60.50,40.00){\vector(1,0){3.00}}
\put(95.50,40.00){\vector(1,0){5.00}}
\put(99.00,30.00){\vector(-1,0){3.00}}
\put(64.00,30.00){\vector(-1,0){3.00}}
\put(49.50,30.00){\vector(-1,0){3.50}}
\put(34.00,30.00){\vector(-4,1){2.00}}
\put(18.00,30.00){\vector(-1,0){2.50}}
\end{picture}
\caption{The graphs contributing to the two point-functions
in the large $N_c$ limit.
a) The class of all strings of constituent quark loops.
The four-fermion vertices are
those in Eq. \protect{\rref{ENJL1}}.
The crosses at both ends are the insertion of the external sources.
b) The one-loop case.}
\rlabel{fig2pt}
\end{center}
\end{figure}
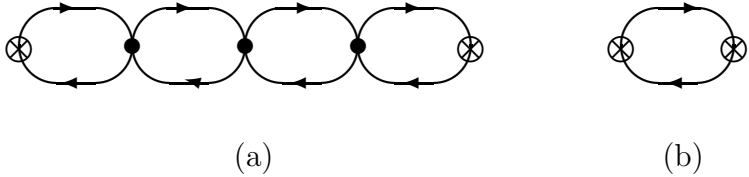
The resummation of strings of constituent
quark bubbles for two-point functions, automatically produces
a pole in all main spin-isospin channels within the purely constituent
quark picture. This is qualitatively the same as the
observed hadronic spectrum.
\item It provides a reasonable description of vector and axial
vector meson phenomenology.
\item Some of the short distance behaviour is even the same as in QCD.
For instance, the Weinberg sum rules \rcite{WE67} are satisfied. This
allows in some cases for a good matching with the short distance behaviour,
see below for an example.
\item The major drawback of the ENJL model is the lack of a
 confinement mechanism.
Although one can always introduce an {\em ad-hoc} confining
potential doing the job.
We smear the consequences of this drawback
by working with internal and external
momenta always Euclidean.
\end{enumerate}

The techniques used here together with more
phenomenological issues are treated
in Refs. \rcite{BBR93,BP94a,BRZ94,PR94}
and reviewed in \rcite{Physrep}.
Some applications to other non-leptonic matrix elements
can be found in \rcite{BdR91} and \rcite{BP95}.
The general conclusion is that within its limitations the
ENJL-type models do capture a reasonable amount of the expected
physics from QCD, its symmetries,
their spontaneous breakdown and even some of its short distance information.

As mentioned before, the Weinberg sum rules \rcite{WE67} are satisfied.
This is a very important
point and is another of the reasons why we have chosen this
model. These relations are needed to obtain
good matching between the low-energy
behaviour and the high-energy one. As an example, they are essential
for the convergence of the hadronic contribution to the
electromagnetic $\pi^+ - \pi^0$ mass difference \rcite{DA67}.
Models to introduce
vector fields like the Hidden Gauge Symmetry (HGS) \rcite{BKY88}
do not have this good intermediate behaviour for some
choices of the parameters. The choice of parameters in the
HGS model used in \rcite{HKS95} to calculate
$a_\mu$ is affected by this problem. For instance,
the contribution to
the above mass difference in the HGS model up to
a cut-off $\nu$, is \rcite{BGun}
\be
\rlabel{HGS}
m_{\pi^+}^2-m_{\pi^0}^2 ={3\over 4} \, {\alpha \over \pi} \, \left[
(1-a)\nu^2-a M_V^2 \log\left(M_V^2/ \nu^2
\right)\right] \, ,
\ee
where $a$ is a parameter of the HGS model.
This obviously diverges badly for $a=2$ which is the
value chosen in \rcite{HKS95}. The same problem is avoided in the ENJL model
\rcite{BRZ94}.

\subsection{The Present Usage.}
\rlabel{3.3}

We will use the ENJL model as a model to fairly describe
in the large $N_c$  limit
 strong interactions between the lowest-lying mesons and,
if needed, external sources.  This is
a tree-level loop model with an explicit cut-off regularization
for one loop parts.
What we mean by a tree-level loop model is the following: a general set
of external sources is connected via full chains, like
the one depicted in Fig. \tref{fig2pt}, to one-loop diagrams
which are also glued through full chains
or four-fermion ENJL vertices.
These are the leading contributions in $1/N_c$.
It is at this level that the hadronic
properties of this model have been tested.
To go beyond this level one would have to include
other operators not
suppressed at the next-to-leading order in $1/N_c$
in the ENJL Lagrangian. At that level one also encounters the problem of
regularizing overlapping divergences in the model.
This is the reason why we shall only apply the ENJL model to
calculate the low-energy large $N_c$ contributions to
$a_\mu^{\rm light-by-light}$.

Since one of the issues that motivated this
calculation was the apparent not fulfilling
of Ward identities
in previous calculations (see comment in \rcite{EI94}),
we want to emphasize here that this model possesses
chiral symmetry and the necessary counterterms are added
so that $n-$point Green functions
fulfil both anomalous and non anomalous Ward
identities \rcite{BP94a,BP94b}.
For instance, the calculation in \rcite{HKS95} assumes ordinary
 VMD for the anomalous sector. It was shown in
\rcite{BP94b} that this VMD breaks the anomalous Ward
identities and one needs for subtractions to restore them.
This is particularly important for the flavour anomaly
contribution to the hadronic light-by-light scattering.
We want to point out also
that in the ENJL model we are using, both anomalous and non anomalous
sectors are described by the same set of parameters. This is not the
case for HGS models where consistency between parameters
in both sectors is not obvious.

\section{Low Energy Large $N_c$ Contributions}
\rlabel{4}
\setcounter{equation}{0}

\begin{figure}
\begin{center}
\leavevmode\epsfxsize=8cm\epsfysize=14cm\epsfbox{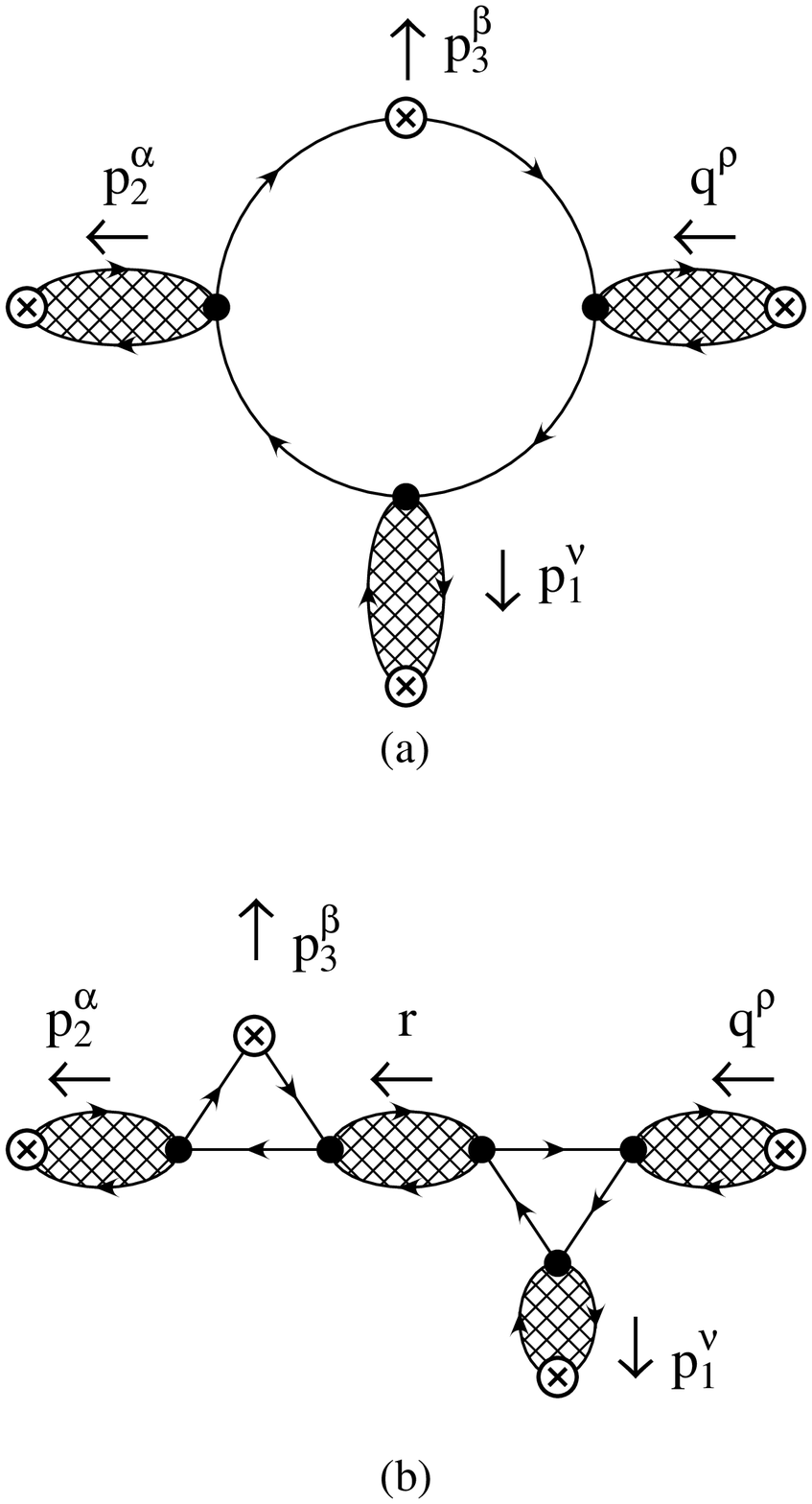}
\end{center}
\caption{The two classes of hadronic light-by-light contributions to
$a_\mu$ at leading ${\cal O}(N_c)$. (a) The four-point functions class.
(b) The product of two three-point functions class.
The dots are ENJL vertices.
The circled crossed vertices are where photons
connect. The cross-hatched loops are full two-point functions
and the lines are constituent quark propagators.}
\label{fig2}
\end{figure}

In this section we discuss the low energy contributions
that appear at large $N_c$ within the framework of the ENJL model
\rcite{deR94}.
For a general contribution, as can be seen in Eqs. \rref{MLB} and
\rref{Damu},
we have to compute the derivative
of the generalized four-point function
$\Pi^{\rho\nu\alpha\beta}(p_1,p_2,p_3)$ in Eq. \rref{fourpoint}
with respect to $p_{3\lambda}$ at $p_3=0$.
The Lorentz structure of this four-point function and some
other technical aspects of its calculation are in Appendix
\tref{A}.
 Since we are dealing with the low energy
contributions to $a_\mu$ we shall only consider the lightest
quark flavours: up, down and strange. Contributions
{}from heavier flavours are discussed in Section \tref{7}.
In the ENJL model there are two classes of contributions to
the generalized four-point function $\fourp$.
The first one is a pure four-point function (see Fig.
   \tref{fig2}(a)).
The second one, which we call three-point-like function
contributions, can be regarded as two
three-point functions glued with one full propagator
(see Fig. \tref{fig2}(b)). Within this framework,
it clearly appears  that these two classes
of contributions to the generalized four-point function
$\fourp$ are different
and therefore should be summed up \rcite{deR94}. The pure
four-point function corresponds
to the so-called quark loop contribution and the three-point-like function
contributions to the meson pole exchange in the language of
Ref. \rcite{HKS95}.
We remind that in our framework both classes of contributions
to $\fourp$ are calculated to all orders in the CHPT expansion.

\subsection{Pure Four-Point Function Contribution}
\rlabel{4.1}

 This contribution is diagrammatically represented
in Fig. \tref{fig2}(a). In the CHPT expansion
the lowest order contribution is order $p^8$, thus potentially
sensitive to the high energy region.
The momenta assignment shown in the figure is the one corresponding to the
first permutation. Whenever we give explicit expressions these are
the ones corresponding to this permutation.
The other five permutations are done analogously.

The pure four-point function is formed by
a one-loop constituent quark vector four-point function
${\overline \Pi}^{\rho\nu\alpha\beta}(p_1,p_2,p_3)$
with the same Lorentz and flavour structure as the four-point function
in Eq. \rref{fourpoint}
with the three vector legs attaching to
the muon line dressed with full vector propagators.
These are the cross-hatched blobs in Fig. \tref{fig2}(b).
The circled crosses vertices
in the figure are where the photon lines connect.
In the large $N_c$ limit there is either a constituent
up, down or strange quark running in the loop.
 For a given quark flavour, this contribution to $\fourp$
can be written using the methods of \rcite{BP94a} as,
\be
\rlabel{PIBAR}
\fourp = {\overline \Pi}_{abcd}
(p_1,p_2,p_3) \, {\cal V}^{abcd\rho\nu\alpha\beta} (p_1,p_2,p_3)
+ \cdots
\ee
with
\ba
\rlabel{vectorprop}
{\cal V}^{abcd\rho\nu\alpha\beta}(p_1,p_2,p_3) &\equiv&
 \left(\frac{g^{a\rho} M_V^2(-q^2)-q^a q^\rho}{M_V^2(-q^2)-q^2}
\right) \,
\left( \frac{g^{b\nu}M_V^2(-p_1^2)-p_1^b p_1^\nu}{M_V^2(-p_1^2)-
p_1^2} \right) \nonumber \\
&\times&  \left(
\frac{g^{c \alpha}M_V^2(-p_2^2)-p_2^c p_2^\alpha}{M_V^2(-p_2^2)-
p_2^2} \right) \, \left(
\frac{g^{d \beta}M_V^2(-p_3^2)-p_3^d p_3^\beta}{M_V^2(-p_3^2)-
p_3^2} \right)\,. \nonumber \\
\ea
Dots stand for other possible contributions to $\fourp$.
The indices $a,b,c,d,e,f$ stand in the remainder for Lorentz indices.
The function $M_V^2(-p^2)$ corresponds to either $\bar u u$,
$\bar d d$ or $\bar s s$ flavour structure and
can be found in Ref. \rcite{BP94a}.
Notice that the full vector propagators in the ENJL model
have the same form as in VMD models but with a
momentum dependent ``vector mass''. See discussion in \rcite{BP94a}
about meson dominance in ENJL models.

After summing over
all the six possible permutations of the external momenta only the
terms proportional to $g_{\mu \nu}$ tensor survive
because of the U(1) gauge invariance. This leads then to
 the phenomenological VMD rule of replacing the
propagator of a photon with momentum $p$ by
\ba
\rlabel{VMDsub}
-\frac{g_{\mu\nu}}{p^2}\, \frac{M_V^2}{M_V^2 - p^2} .
\ea
Thus implying that this rule  preserves the Ward identities
after summing over all the permutations.

We compute the one-constituent quark loop
following the analysis in Appendix \tref{A} and using a proper
time regulator with a physical cut-off $\Lambda$.

The contribution of
${\overline \Pi}^{abcd}(p_1,p_2,p_3)$ in eq. \rref{PIBAR}
to $a_\mu$ can be decomposed in terms of 32 independent amplitudes
(see App. \tref{A}).
After integrating over
the momentum running in the loop, these amplitudes will be integrals
over the three Feynman parameters introduced by the standard procedure
of reducing the four internal propagators to a power of one
propagator. Moreover, after taking the derivative with respect
to $p_{3\lambda}$ at $p_3=0$ one of the Feynman parameters can
be integrated out analytically.
The general form of  each one of the 32 functions in
${\overline \Pi}^{abcd}(p_1,p_2,p_3)$ for
a constituent quark of flavour $i$ contributing
to $a_{\mu}^{\rm light-by-light}$ is thus
\be
\rlabel{gener}
\frac{N_c}{16\pi^2}\, Q_i^4  \,
{\dis \int^1_0} {\rm d}x
{\dis \int^{1-x}_0}{\rm d}y \,
\frac{f(x,y)}{M^4(x,y)} \Gamma_2(M^2(x,y)/ \Lambda^2),
\ee
where
\be
M^2(x,y)\equiv M_i^2 - x(1-x)(p_1+p_2)^2 -y(1-y)p_2^2+
2xy(p_1+p_2)\cdot p_2 ,
\ee
and
\be
\Gamma_2(\epsilon) = (1+\epsilon)
e^{-\epsilon} \, .
\ee
Here $M_i$ is the constituent quark mass with flavour $i$.
The Feynman parameters $x$ and $y$, together with
 the other five degrees of
 freedom for the two external muon loops  in Eq. \rref{MLB},
produce a seven dimensional integral that we perform using
the Monte Carlo routine VEGAS.
The numerical results of this contribution will be discussed
in Section \tref{6}.

As a check we have reproduced the results for the constituent
quark and muon loops in \rcite{KNO85}
and the electron loop result in \rcite{LS77}. This is the calculation
using \rref{PIBAR} with
${\cal V}^{abcd\rho\nu\alpha\beta} = g^{a\rho}g^{b\nu}g^{c\alpha}g^{d\beta}$.
We have also re-scaled the relevant masses. For the constituent quark mass
we have used $M_Q=$ 300 MeV. The results are in Table
\tref{tablex}. The numbers in brackets are
the VEGAS quoted errors. The last row gives the known results.
\begin{table}
\begin{center}
\begin{tabular}{|c|c|c|c|}
\hline
Cut-off & $a_\mu \times 10^{7}$ &
$a_\mu \times 10^{9}$ &$a_\mu\times 10^{9}$ \\
$\mu$ & Electron & Muon & Constituent Quark \\
(GeV)  &  Loop & Loop & Loop \\
\hline

0.5 &  2.41(8)   & 2.41(3)  & 0.395(4)  \\
0.7 &  2.60(10)   & 3.09(7)  & 0.705(9)  \\
1.0 & 2.59(7) & 3.76(9)& 1.10(2)\\
2.0 & 2.60(6) & 4.54(9)& 1.81(5)\\
4.0 & 2.75(9)& 4.60(11) & 2.27(7)\\
8.0 & 2.57(6)& 4.84(13) & 2.58(7)\\
\hline
Known Results & 2.6252(4)& 4.65 & 2.37(16) \\
\hline
\end{tabular}
\caption{The results from a pure fermion loop compared
with known numerical and analytical results. The known result for the
electron loop has been partially analytically evaluated in
\protect{\rcite{LS77}}, while the quoted values for the muon loop and the
constituent quark loop are respectively analytical and numerical results
of \protect{\rcite{KNO85}}.}
\end{center}
\label{tablex}
\end{table}
Within the numerical uncertainty of our calculation the agreement is good.
It should be remarked that there are cancellations present at high energy.
These contributions make the estimate numerically more uncertain for
higher values of the cut-off $\mu$. As can be seen the
electron loop has essentially reached its final value at a cut-off of 500~MeV.
The same is not true for the muon loop or the constituent quark loop.
The muon loop reaches its asymptotic
value essentially at a cut-off of about 2~GeV. The constituent quark loop
reaches its asymptotic value above 4~GeV. We have also calculated the
same quantities for a few
higher cut-offs and observed their stabilization within the numerical error.
These sample calculations show that a proper study of the cut-off
behaviour for the more difficult to estimate
complete hadronic contribution is definitely necessary.

\subsection{Three-Point-Like Function Contributions}
\rlabel{4.2}

 This class of contributions is diagrammatically represented
in Fig. \tref{fig2}(b). There are two permutations
of the vector legs for each of the two three-point functions.
In addition there are three possible sets of two momenta
out of the four external vector legs momenta. This makes
twelve possible permutations
to be considered for three-point-like
function contributions. The momenta shown in the
figure are the ones corresponding to the first permutation.
Whenever we give explicit expressions these are
the ones corresponding to this permutation.
The other eleven permutations are done analogously.

In this case we have two one-loop  three-point functions
with two vector legs each one
glued with a full two-point function
that can be either pseudoscalar, scalar,
mixed pseudoscalar--axial-vector, or  axial-vector.
For intermediate vector two-point functions the result
is zero because of Furry's theorem.
Three of the vector legs here are then attached to the muon line with
dressed full vector propagators just as in the case of the pure
four-point function discussed in the previous section.
Again, in the large $N_c$ limit,
there is either an up, down or strange constituent quark running in the loop.
Technically we have used two different approaches to calculate
this type of contributions. One is using the Ward
identities  for the four-point function $\fourp$. In this way
one has to determine the 32 amplitudes of $\fourp$
contributing to $a_\mu^{\rm light-by-light}$, see
Appendix \tref{A}. The other way is constructing
explicitly the complete $\fourp$ starting from the three- and
two-point functions. Here one relies on the Ward identities
for three-point functions. As a check we verified that both ways
agree exactly.

In what follows we study each type of exchange (scalar, pseudoscalar
and axial-vector) separately.

As an additional numerical check we have  also calculated the
pion exchange contribution with vector meson dominance
of the rho meson in the vector legs and when
the anomalous vertices are from the order
$p^4$ Wess-Zumino effective action. Our result agrees exactly with
that in Eq. (4.1) of \rcite{HKS95b} which quotes $-55.60(3)\cdot
10^{-11}$.

The two-point function involves the integration over one
Feynman parameter and each one of the two three-point
functions the integration over two more
Feynman parameters. These integrals have been evaluated using
Gaussian integration. To obtain $a_\mu$, one has to convolute
these three-point-like contributions with the five dimensional
space integral of the external two muon-loops
in Eq. \rref{MLB} which  has been performed using
the Monte Carlo routine VEGAS.
The numerical results for these contributions will be discussed
in Section \tref{6}.

\subsubsection{Scalar Exchange}
\rlabel{4.2.1}

For the scalar exchange, the
lowest order contribution in the CHPT expansion is order
$p^8$. This contribution to $\fourp$ can be written
as
\ba
\rlabel{threescalar}
\fourp &=&  \! {\overline \Pi}_{ab}^{VVS}(p_1,r)
g_S \left(1+g_S
\Pi^S(r)\right) {\overline \Pi}_{cd}^{SVV}(p_2,p_3)
{\cal V}^{abcd\rho\nu\alpha\beta}
(p_1,p_2,p_3)\nonumber\\&&+\cdots ,\nonumber \\
\ea
where ${\cal V}^{abcd\rho\nu\alpha\beta} (p_1,p_2,p_3)$
has been defined in Eq. \rref{vectorprop} and $r=p_2+p_3$.
The full two-point function $\Pi^S(p)$ can be found in
Ref. \rcite{BP94b} and the one-loop three-point functions
${\overline \Pi}^{VVS}_{\mu\nu}(p,q)$ and
${\overline \Pi}^{SVV}_{\mu\nu}(p,q)$ are in Appendix \tref{B}.
Eq. \rref{threescalar} can be understood as follows: $g_S ( 1+g_S
\Pi^S(r) ) $ is the scalar propagator connecting the two vertices
${\overline \Pi}_{cd}^{SVV}(p_2,p_3)$ and ${\overline \Pi}_{ab}^{VVS}(p_1,r)$.
The whole diagram is then glued to the external photon lines with the
vector propagators given by ${\cal V}^{abcd\rho\nu\alpha\beta}$.

\subsubsection{Pseudoscalar Exchange}
\rlabel{4.2.2}

For the pseudoscalar exchange
the lowest order contribution in the CHPT expansion is
order $p^6$.
This, together with the fact that it involves two flavour anomalous
vertices points out that this contribution
could be the leading one. Considering as
part of the pseudoscalar exchange all those terms proportional
to a pseudoscalar
propagator, this contribution also includes the
pseudoscalar--axial-vector mixed terms. Its expression is given by
\ba
\rlabel{pseudoex}
\fourp &=& \left[{\overline \Pi}_{ab}^{VVP}(p_1,r) \left( 1+g_S
\Pi^P(r) \right) {\overline \Pi}_{cd}^{PVV}(p_2,p_3)
\right. \nonumber \\
&&- g_V {\overline \Pi}_{ab}^{VVP}(p_1,r)
\Pi^{P\mu}(r)  {\overline \Pi}_{\mu cd}^{AVV}(p_2,p_3)
\nonumber \\
&&- g_V \left. {\overline \Pi}_{ab\mu}^{VVA}(p_1,r)
\Pi^{P\mu}(-r)  {\overline \Pi}_{cd}^{PVV}(p_2,p_3)
\right] \nonumber \\ &\times& g_S
{\cal V}^{abcd\rho\nu\alpha\beta} (p_1,p_2,p_3)+\cdots
\ea
where ${\cal V}^{abcd\rho\nu\alpha\beta}(p_1,p_2,p_3)$
has been defined in Eq. \rref{vectorprop} and $r=p_2+p_3$.
The full two-point functions $\Pi^P(p)$ and $\Pi^P_\mu(p)$
 and the one-loop three-point functions
${\overline \Pi}^{PVV}_{\mu\nu}(p,q)$ and
${\overline \Pi}^{VVP}_{\mu\nu}(p,q)$
can be found in Ref. \rcite{BP94a}.
See Eq. \rref{BARME} in
Appendix \tref{B} for the explicit expression.
The mixed two-point function $\Pi^P_\mu(p)$ can
be written as \rcite{BP94a},
\be
\Pi^P_\mu (p) \equiv i p_\mu \, \Pi^P_M(p^2) .
\ee
Since both the $VVA$ and the $AVV$ one-loop three-point
functions are multiplied by $r_\mu$,
we use the one-loop anomalous Ward
identity in Eq. (4.24) of \rcite{BP94a}
and the prescription given in \rcite{BP94b}
to rewrite Eq. \rref{pseudoex} as
\ba
\rlabel{pseudoex2}
\fourp& =& g_S {\overline \Pi}_{ab}^{VVP} (p_1,r)
\left( 1+g_S \Pi^P(r) -
4 g_V M_i \Pi^P_M(r^2) \right) \nonumber \\
&&\times {\overline \Pi}_{cd}^{PVV}(p_2,p_3) \,
{\cal V}^{abcd\rho\nu\alpha\beta}(p_1,p_2,p_3) \nonumber \\
&+& 2 M_Q g_S g_V \Pi^P_M(r^2)  \left[
{\overline \Pi}_{ab}^{VVP}(p_1,r)
\left\{ {\overline \Pi}_{cd}^{PVV}(p_2,p_3)
{\left. \right|^{M_i=M_Q}_{p_2^2=p_3^2=r^2=0}} \right\}
\right. \nonumber \\ &\times&
 \left( \frac{g^{a\rho} M_V^2(-q^2)-q^a q^\rho}
{M_V^2(-q^2)-q^2} \right) \,
\left( \frac{g^{b\nu} M_V^2(-p_1^2)-p_1^c p_1^\nu}
{M_V^2(-p_1^2)-p_1^2} \right) g^{c\alpha} g^{d\beta}\,
\nonumber \\ &+&
 \left\{ {\overline \Pi}^{ab}_{VVP}(p_1,r)
{\left. \right|^{M_i=M_Q}_{p_1^2=r^2=q^2=0}} \right\}
{\overline \Pi}^{cd}_{PVV}(p_2,p_3) \nonumber \\
&\times& \left.   g^{a\rho} g^{b\nu}
\left( \frac{ g^{c\alpha}M_V^2(-p_2^2) -p_2^c p_2^\alpha}
{M_V^2(-p_2^2)-p_2^2} \right)
\left( \frac{g^{d\beta}M_V^2(-p_3^2)- p_3^d p_3^\beta}
{M_V^2(-p_3^2)-p_3^2} \right) \right]
\nonumber \\ &+& \cdots \, ,
\ea
where $M_i$ is the constituent quark mass for the quark with flavour $i$
and $M_Q$ its value in the chiral limit.
The first two lines above come from applying naive Ward
identities to the
axial-vector leg, while the remaining contributions
are the subtractions needed
to fulfil the anomalous Ward identities. In addition,
${\overline  \Pi}^{PVV}_{\mu\nu} (p,q)$  contains subtractions
also determined by the anomalous Ward identities \rcite{BP94a}.
See  Eq. \rref{anomform} in Appendix \tref{B}.
The expression \rref{pseudoex2} also shows that
the pseudoscalar exchange contribution always contains
at least one vector meson propagator.
 Due to this fact and the presence of the subtractions in \rref{pseudoex2},
the $\pi^0\gamma^*\gamma^*$ vertex
goes to a constant when the  vector legs' Euclidean momenta are
very large. Therefore,
although this contribution when summed over all possible
permutations is convergent by itself  because
of gauge invariance (see Section \tref{2}),
the subtraction terms make it very slowly convergent. We know that
in QCD the $\pi^0\gamma^*\gamma^*$
vertex goes like $1/Q^2$ at large Euclidean momentum \rcite{GL95}.
This behaviour is also supported by the measured $\pi^0 \gamma
\gamma^*$ form factor in the Euclidean region at CELLO \rcite{CELLO}
and CLEO-II \rcite{CLEO} detectors.
This indicates again that although this  model gives the right
contribution for energies below or
around $\Lambda$, it breaks down above. We shall  estimate the
intermediate and high energy region contributions for
the pseudoscalar exchange in Section \tref{6}.

Although the present section is devoted to the large $N_c$ contributions,
it is worth to discuss the main $1/N_c$ corrections to the
 pseudoscalar exchange at this point. These are the effects
of the U(1)$_A$ anomaly and were already included in the Erratum in
Ref.  \rcite{BPP95}. In the chiral limit and in the
large $N_c$ limit there are nine pseudo-Goldstone bosons
\rcite{WI79}: $\pi^\pm$, $\pi^0$, $K^\pm$,
$K^0$, $\overline K^0$, $\eta_8$ and $\eta_1$.
Under flavour SU(3) they transform as a nonet
multiplet. However nonet symmetry is broken by $1/N_c$ effects
due to the U(1)$_A$ anomaly. These effects cause
the isospin zero mass eigenstates to become the $\eta$ and
$\eta'$ states. They  also increase the mass of the
$\eta'$ meson to 958 MeV. These $1/N_c$ corrections are thus quite
relevant to the pseudoscalar exchange. We have taken them into
account by using the physical $\pi^0$, $\eta$ and $\eta'$
mass eigenstates as propagating states. This already gives
the bulk of the effects of the U(1)$_A$ anomaly.
Higher order corrections are negligible and within
the quoted error. The results of using either
$\bar u u$, $\bar d d$, and $\bar s s$ basis of states
for the large $N_c$ limit or the physical $\pi^0$, $\eta$ and $\eta'$
basis are given in Section \tref{6}.

\subsubsection{Axial-Vector Exchange}
\rlabel{4.2.3}

For the axial-vector exchange, the
lowest order contribution in the CHPT expansion is order $p^8$.
This contribution to $\fourp$ can be written
as
\ba
\rlabel{threeaxial}
\fourp& =& - g_V \, {\overline \Pi}_{abe}^{VVA}(p_1,r) \left(
g^{ef}-g_V \Pi_A^{ef}(r) \right)
{\overline \Pi}_{fcd}^{AVV}(p_2,p_3)
\nonumber\\&&\times{\cal V}^{abcd\rho\nu\alpha\beta} (p_1,p_2,p_3)+ \cdots
\ea
where ${\cal V}^{abcd\rho\nu\alpha\beta} (p_1,p_2,p_3)$
has been defined in Eq. \rref{vectorprop} and $r=p_2+p_3$.
The full axial-vector two-point function $\Pi_A^{\mu\nu}(p)$
can be found in Ref. \rcite{BP94a}
and the one-loop three-point functions
${\overline \Pi}^{VVA}_{\mu\nu\alpha}(p,q)$ and
${\overline \Pi}^{AVV}_{\mu\nu\alpha}(p,q)$
are in Appendix \tref{B}.

Using the one-loop anomalous Ward identity in Eq. (4.24) of
 \rcite{BP94a} and the prescription given in \rcite{BP94b}
for the terms in \rref{threeaxial}
where $r_\mu$ multiplies either the $VVA$ or the $AVV$
three-point functions, we can rewrite  \rref{threeaxial} as
\ba
\rlabel{threeaxial2}
\fourp& =& - g_V \, {\overline \Pi}_{abe}^{VVA}(p_1,r)
 g^{ef} \left( 1 + g_V r^2 \Pi_A^{(1)}(-r^2) \right)
{\overline \Pi}_{fcd}^{AVV}(p_2,p_3)
\nonumber\\&&\times{\cal V}^{abcd\rho\nu\alpha\beta} (p_1,p_2,p_3)
- 2 g_V^2 \left( \Pi_A^{(0)}(-r^2) + \Pi_A^{(1)}(-r^2) \right)
\nonumber \\ &\times&
\left\{ \left( \frac{g^{a\rho} M_V^2(-q^2)-q^a q^\rho}
{M_V^2(-q^2)-q^2} \right) \,
\left( \frac{g^{b\nu} M_V^2(-p_1^2)-p_1^c p_1^\nu}
{M_V^2(-p_1^2)-p_1^2} \right) \right. \nonumber \\
&& \left. M_i {\overline \Pi}_{ab}^{VVP}(p_1,r)
- M_Q  {\overline \Pi}^{\rho \nu}_{VVP}(p_1,r)
{\left. \right|^{M_i=M_Q}_{p_1^2=r^2=q^2=0}} \, \right\} \nonumber \\
&\times& \left\{ \left( \frac{ g^{c\alpha}M_V^2(-p_2^2)- p_2^c p_2^\alpha}
{M_V^2(-p_2^2)-p_2^2} \right) \,
\left( \frac{g^{d\beta}M_V^2(-p_3^2)- p_3^d p_3^\beta}
{M_V^2(-p_3^2)-p_3^2} \right) \right. \nonumber \\
&& \left. M_i {\overline \Pi}_{cd}^{PVV}(p_2,p_3)
 - M_Q  {\overline \Pi}^{\alpha \beta}_{PVV}(p_2,p_3)
{\left. \right|^{M_i=M_Q}_{p_2^2=p_3^2=r^2=0}}  \right\}
\nonumber \\ &+& \cdots \, .
\ea
The expression for the two-point axial amplitudes
$\Pi_A^{(0)}(-r^2)$ and $\Pi_A^{(1)}(-r^2)$
can be found in \rcite{BP94a}.
The kinematical pole of the transverse part at $r^2=0$
disappears in the combinations $1+g_V r^2 \Pi_A^{(1)}(-r^2)$ and
$\Pi_A^{(0)}(-r^2)+\Pi_A^{(1)}(-r^2)$.
The last combination contains a pseudoscalar
propagator which is often included in the pseudoscalar exchange
contribution.

\section{Low Energy Next-to-Leading in $1/N_c$
Contributions}
\rlabel{5}
\setcounter{equation}{0}

In this section we discuss the low energy
next-to-leading in $1/N_c$ contributions to
$a_\mu^{\rm light-by-light}$. In the previous
sections we have discussed the low energy
large $N_c$ contributions in the context of the ENJL model.
We included the effects of the U(1)$_A$ anomaly which
are next-to-leading in $1/N_c$ as well. Here we address
the calculation of  the other
${\cal O}(1)$ corrections in the $1/N_c$ expansion.
In the language of the ENJL model,
these corrections are given by loops of strings of quark bubbles.
They contain one closed loop of a string of bubbles like the one
in Fig. \tref{fig2pt}(a).
 This loop is then connected in all possible ways
to the photons using strings of bubbles. In a mesonic picture they correspond
to one meson loop contributions.
This meson loop can then substitute any one-loop constituent quark
(bubble) in diagrams in Figs. \tref{fig2}(a) and \tref{fig2}(b).
We end up with two classes of $O(1)$ contributions.
The $1/N_c$ next-to-leading corrections to the full two-point
functions, i.e. to meson
masses and couplings and the $1/N_c$ next-to-leading
corrections to the three- and
four-point one-loop functions, i.e. to vertices.
In fact, the major $1/N_c$ correction
to pseudoscalar two-point functions, namely the
U(1)$_A$ anomaly, was already estimated in Section
\tref{4}. The other $1/N_c$ corrections to two-point functions
should be small since
the phenomenological analysis in
Refs. \rcite{BBR93,BRZ94} and \rcite{BP94a} fits very well.
We thus expect these to be already included in the error of the model for
 the large $N_c$ results.
Therefore, we shall only consider here
the $1/N_c$ corrections to the vertices.

Unfortunately,
at present, these type of contributions cannot be fully
treated in the ENJL model. Some of the reasons
were given in Section \tref{3}. In its
present form the ENJL model we are using is just well defined
in the large $N_c$ limit. However, the fact that these type of
contributions  are of different  large $N_c$ counting with respect to the
ones treated in Section \tref{4},
permits the separate treatment we follow below for them.

Four-point functions can be also calculated
at very low energy within CHPT \rcite{WE79}. In this regime
the relevant degrees of freedom are the
lowest pseudoscalar mesons, while vector, axial-vector and scalar
resonances have been integrated out.
Their effects are included in the couplings of the
CHPT Lagrangian \rcite{EGPR}.
CHPT becomes then a good tool to study  strong interactions of
the lowest pseudoscalar mesons with external sources.
In this framework we need both $\gamma^* P^+ P^-$ and
$\gamma^* \gamma^* P^+ P^-$ vertices, where $P$ is pion or
kaon.\footnote{Contributions from vertices with more photons start at
 higher order in CHPT and are therefore suppressed with respect to
the contributions of  $\gamma^* P^+ P^-$ and $\gamma^* \gamma^* P^+ P^-$.}
The first vertex is well known phenomenologically and
VMD models give a very good description of it. On the
contrary, not much is known phenomenologically about the
second one. This fact induces a large model dependence since one
can construct many models satisfying the relevant Ward
identities and  with different degrees of VMD. One can use
for instance the HGS model as in \rcite{HKS95} where there is
no complete VMD for the $\gamma^* \gamma^* P^+ P^-$ vertex.
We shall discuss this contribution in a complete VMD model both for
$\gamma^* P^+ P^-$ and $\gamma^* \gamma^* P^+ P^-$ vertices.
This is done inspired by the form of the ${\cal O} (N_c)$
contributions in the ENJL model.

In previous sections we have seen  that the most general expression
for the four-point function
$\fourp$ at ${\cal O}(N_c)$ in the ENJL model contains
the one-loop four- or three-point-like
functions, which give the lowest order in the CHPT counting,
multiplied by the ENJL vector meson propagators in Eq. \rref{vectorprop}.
Inspired by this behaviour
we  saturate the ${\cal O}(1)$ contribution by
one loop of charged pion or kaon mesons using lowest
order CHPT photon-pion vertices, i.e. ${\cal O}(p^2)$,
 multiplied by
the ENJL vector meson propagators in \rref{vectorprop}.
The main difference with a full ENJL calculation
here is that we substitute the momentum dependent pion mass
and coupling by their experimental values.

To understand the sensitivity to the momenta dependence of the
vector meson propagators we have numerically studied the difference
between the case with vector propagators containing a constant vector
mass and the case with a momentum dependent one.
Each choice gives different high energy behaviour
of the photon-pion vertices.

For the  pseudo-Goldstone bosons loops contribution to $\fourp$,
the lowest order in the CHPT counting is order $p^4$.
This also points out
that this contribution could be dominated by lower energy
regions than the other contributions analyzed in previous sections
which started at order $p^6$.
Since we are including as propagating states only the lowest
pseudoscalar mesons and the rho vector, our approach
will be only valid
for energies below or around 1 GeV. Above this energy axial-vector mesons
become also dynamical states. The saturation of this contribution
{}from physics at scales around  1 GeV can only be confirmed
{\em a posteriori}. The numerical results concerning this
contribution are presented in Section \tref{6}.

We now proceed to analyze in more detail the two types
of possible contributions:
the pure four-point function and three-point-like function
contributions.  In particular,
the three-point-like contributions
can be seen as the diagram in Figure \tref{fig2}(b)
where one or both one-constituent-quark loop
three-point functions are replaced with
charged pion or kaon loops.
Due to parity, the intermediate two-point function glueing
the two three-point functions can be either vector
or scalar. The vector contribution is again zero because
of Furry's theorem (this can be  better verified in the ENJL inspired
diagrams where the vector legs couple to fermion lines).
The scalar contribution we expect to be very much suppressed
like in the ${\cal O}(N_c)$ case (see
numerical results in Section \tref{6}). The important point
here is that there are no anomalous contributions since
now we have mesons running in the three-point
functions. This makes this contribution to be in the range of
the expected CHPT counting and not anomalously large as
we obtained for the ${\cal O}(N_c)$ pseudoscalar exchange.

Therefore we only estimate the dominant pure four-point function
contribution to $\fourp$ in Eq.  \rref{fourpoint}.
This can be seen as the diagram in Fig. \tref{fig2}(a) where
the one-loop constituent quark four-point function is now a loop of
pseudoscalar mesons.  It can be written as
\be
\rlabel{pionloop}
\fourp ={\overline \Pi}_{abcd}
(p_1,p_2,p_3) \, {\cal V}^{abcd\rho\nu\alpha\beta}(p_1,p_2,p_3)
+\cdots
\ee
where ${\overline \Pi}_{abcd} (p_1,p_2,p_3)$
is the ${\cal O}(1)$ in $1/N_c$
contribution from charged pion and kaon loops using
lowest order in CHPT photon-pion vertices. For that, we compute
at lowest order in CHPT, the quark vector
current appearing in the definition of $\fourp$ \rcite{WE79}
\ba
V^\mu_a (x) \Rightarrow
i Q_a \left( \left[ \pi^+ (x), D^\mu \pi^- (x) \right] +
\left[ K^+ (x), D^\mu K^- (x) \right] \right)_{aa} \, ,
\ea
where the subscript $aa$ means that we take the $aa$ component
in flavour space. Due to the covariant derivative
$D_\mu P \equiv \partial_\mu P - i |e| [A_\mu, P]$,
the term above gives rise both to $P^+ P^- \gamma^*$ and
$P^+ P^- \gamma^* \gamma^*$ vertices. Of course, the full contribution
to $\fourp$ at a given  order in CHPT, in this
 case ${\cal O}(p^2)$, has to be gauge covariant.
We construct the one-loop four-point function
${\overline \Pi}_{abcd} (p_1,p_2,p_3)$  following again the
analysis in Appendix \tref{A}, where we need to determine
 the 32 independent amplitudes that
contribute to the  $a^{\rm light-by-light}_\mu$.
These amplitudes have  always the Lorentz indices saturated
by external momenta indices. Notice that vertices with one photon
are proportional to external momenta, while vertices
 with two photons are proportional
to  $g_{\mu\nu}$ tensors. Thus we only need to
compute the UV convergent amplitudes and reconstruct
 the contribution from the two-photons--two-mesons
vertices by using gauge invariance.
The full meson one-loop with order $p^2$ vertices
explicitly satisfies the chiral and U(1) gauge Ward identities.
In fact the amplitudes
one gets for the 32 independent functions are very similar
to the ones found in the one-loop constituent quark
amplitudes in Eq. \rref{gener}. The general form we obtain
for them  is
\be
\rlabel{generpi}
-\frac{1}{16\pi^2} \left(Q_u-Q_d \right)^4
{\dis \int^1_0} {\rm d}x
{\dis \int^{1-x}_0}{\rm d}y \,
\frac{g(x,y)}{\tilde M^4(x,y)} ,
\ee
with
\be
\tilde M^2(x,y)\equiv m_P^2 - x(1-x)(p_1+p_2)^2 -y(1-y)p_2^2+
2xy(p_1+p_2)\cdot p_2
\ee
and $m_P$ is the mass of the pseudoscalar meson $P$.
The one-loop meson pure four-point function is multiplied
with the propagator in \rref{vectorprop} to give the full
four-point function of Eq. \rref{pionloop}.

If the one-loop four-point
function satisfies the Ward identities, as it does, also the full
four-point function \rref{pionloop} will satisfy them since
\be
q_\mu \, \left(g^{\mu\nu} M_V^2 - q^\mu q^\nu \right)
=q^\nu \left(M_V^2 - q^2 \right) \, ,
\ee
independently of whether $M_V$ is momentum dependent or not.
As we already said  for the
one-loop constituent quark loop contribution,
after summing over all possible
permutations of the three vector legs,
only the terms proportional to $g_{\mu\nu}$ in
\rref{vectorprop} survive since the meson
one-loop four-point function satisfies the Ward identities.
This leaves the phenomenologically VMD rule in
Eq. \rref{VMDsub} to work here too\footnote{In fact,
the complete VMD we are using is identical
to the so-called naive VMD model in Ref. \rcite{KNO85,HKS95}
which therefore does not break either the chiral Ward identities
or the U(1) electromagnetic covariance.}.
This eliminates the worries about the
fulfilling of chiral symmetry when
using this substitution \rcite{EI94,HKS95}.

That this procedure is fully chiral and U(1) gauge invariant can be seen
simply by constructing a Lagrangian  with full electromagnetic gauge
and chiral invariance and that has complete VMD for both
$\gamma P^+ P^-$ and $\gamma \gamma P^+ P^-$ vertices.
The following Lagrangian contains couplings of pions and
photons to all orders in external momenta and reproduces
the full VMD amplitude without inducing any extra vertices
of photons and pseudoscalar mesons.
\ba
\rlabel{VMDLAG}
{\cal L}_{\mbox{VMD}}&=&\frac{f_\pi^2}{4}
\tr \left( {\cal D}_\mu U {\cal D}^\mu U^\dagger +\chi U^\dagger
+U\chi^\dagger\right)\nonumber
\\ &+&
\sum_{n\geq 0} i a_n
\tr\left[\left( {\cal D}^{2n}
{\cal D}^\mu L_{\mu\nu}\right) U^\dagger {\cal D}^\nu U
+\left({\cal D}^{2n} {\cal D}^\mu R_{\mu\nu}\right) U
{\cal D}^\nu U^\dagger\right] \nonumber \\ &+&
\sum_{n,m\geq 0} b_{nm} \tr\left[\left({\cal D}^{2n}{\cal D}_\mu
L^{\mu\nu}\right)U^\dagger
\left({\cal D}^{2m}{\cal D}^\alpha R_{\alpha\nu}\right)U\right].
\ea
The $3 \times 3$ flavour
matrix $U$ contains the pseudoscalar meson fields \rcite{WE79} and
\be
{\cal D}_\mu U =
\partial_\mu U -i(v_\mu +a_\mu) U
+i U (v_\mu - a_\mu) \,.
\ee
Here $v_\mu$, $a_\mu$  are external vector and  axial-vector fields,
while the field $\chi = 2~B_0{\cal M}+...$ contains the current
quark mass matrix ${\cal M}$.
The photon field is contained in $v_\mu$.

The field strengths $R(L)_{\mu\nu}$ are constructed out of the
fields $r(l)_\mu = v_\mu +(-) a_\mu$.
The covariant derivatives act on $R(L)_{\mu \nu}$ as follows
\be
{\cal D}_\alpha R(L)_{\mu\nu} =
\partial_\alpha R(L)_{\mu\nu} -i[r(l)_\alpha, R(L)_{\mu\nu}].
\ee
The form of the terms in \rref{VMDLAG}
has been chosen such that there are no vertices with three or more photons
 interacting with pions generated.
The first line is the lowest order CHPT Lagrangian.
The second line contains one- or two-photons couplings
to pseudoscalar mesons
while the last line  contains only two-photons couplings
to pseudoscalar mesons.

The vertex for a charged pion with incoming momentum $p$  and a photon with
outgoing momentum $k$ and polarization  $\epsilon^\mu$ is given by
\ba
&&i|e| \left(2p-k\right)^\nu
\left\{ g_{\mu\nu}+ \left(k^2 g_{\mu\nu}-k_\mu k_\nu\right)
\frac{2}{f_\pi^2}
\sum_{n\geq 0} a_n \left(-k^2\right)^n \right\}\,.
\ea
The first term is the lowest order vertex. With the choice
\be
a_n = -\frac{f_\pi^2}{2} \left(\frac{-1}{M_V^2}\right)^{n+1}
\ee
this reproduces the phenomenological
complete VMD behaviour for the $\gamma^* P^+ P^-$
vertex. But notice that any VMD-like $M_V^2(k^2)/(M_V^2(k^2)-k^2)$
behaviour, with $M_V^2(0) \neq 0$, as the one obtained in
the ENJL model can be reproduced by choosing
 the $a_n$ appropriately. I.e.
\be
 \frac{2}{f_\pi^2} \sum_{n\geq 0} a_n \left(- k^2 \right)^n=
 \frac{1}{M_V^2(k^2)-k^2}.
\ee

We now turn to the $\gamma^* \gamma^* P^+ P^-$ vertex.
If  the two photons outgoing  momenta  are $k_1$ and $k_2$ and
their polarizations  $\epsilon^\mu$ and $\epsilon^\nu$ respectively,
that vertex is given by
\ba
&&2ie^2 g_{\mu\nu} \nonumber \\
&&+4i \frac{e^2}{f_\pi^2}\,
 \left(k_1^2 g_{\mu\nu}-k_{1\mu} k_{1\mu}\right)
\sum_{n\geq 0} a_n\left(-k_1^2\right)^n \nonumber \\
&&+4 i\frac{e^2}{f_\pi^2}\,
 \left(k_2^2 g_{\mu\nu}-k_{2\mu} k_{2\nu}\right)
\sum_{n\geq 0} a_n\left(-k_2^2\right)^n
\nonumber\\
&&- 2 i\frac{e^2}{f_\pi^2}\,
\left(k_1^2 g_\mu^{.\alpha}-k_{1\mu} k_1^\alpha \right)
\left(k_2^2 g_{\alpha\nu}-k_{2\alpha} k_{2\nu}\right)
\nonumber \\
&\times&
\sum_{n,m\geq 0}b_{nm}
\left[\left(-k_1^2\right)^n\left(-k_2^2\right)^m+
\left(-k_2^2\right)^n\left(-k_1^2\right)^m\right] \,.
\ea
Here we see that,  keeping gauge and chiral invariance fully
satisfied, this vertex is rather unconstrained. The choice $b_{nm}=0$
reproduces the HGS model used in \rcite{HKS95,HKS95b} with $a=2$.
The difference between that model and the complete VMD model
only  starts at order $p^6$ in the chiral counting.

A large number of other choices are however possible.
In particular the choice
\be
b_{nm} = -\frac{2}{f_\pi^2} \, a_n \, a_m
\ee
reproduces the complete VMD amplitude for $\gamma^* \gamma^* P^+ P^-$
used in this work. Again depending on the coefficients $a_n$,
one can have the ENJL vector meson propagator or any other one like
the phenomenological complete VMD mentioned above.
For comparison we use both, see Section \tref{6}.
It is also possible to
add chiral invariant terms that will produce a direct dependence
on $k_1\cdot k_2$.
This last possibility is realized by adding terms like
\be
\left({\cal D}^{2n} {\cal D}^\beta {\cal D}^\mu L_{\mu\nu}\right)
U^\dagger \left({\cal D}^{2m}{\cal D}_\beta
 {\cal D}^\alpha R_{\alpha\nu}\right) U.
\ee

The numerical results  for the pseudoscalar mesons loops contribution
are discussed in Section \tref{6}.

\section{Numerical Results}
\rlabel{6}
\setcounter{equation}{0}

In this section we give the numerical results for the
low energy calculation of the hadronic light-by-light
contributions to $a_\mu$ presented in Sections \tref{4} and
\tref{5}.

We first analyze the result for the
large $N_c$ limit calculation including the effects of the
U(1)$_A$ anomaly as explained in Section
\tref{4}. Since we are dealing with a low-energy model,
as mentioned before,
it is necessary to study the dependence on a high-energy cut-off $\mu$
on the vector legs' momenta.

For the seven dimensional integral of the
pure four-point function (or one constituent
quark loop) contribution, we used a statistics of 20 iterations
with $10^5$ points  in the Monte Carlo
routine VEGAS, while for the
two muon loops five dimensional
integral of the three-point-like function contributions we used a
statistics of 20 iterations with 5000 points
in the same Monte Carlo routine. This statistics is equivalent
to the one used in the seven dimensional integral case.
For the  two- and three-point functions needed in these
three-point-like contributions we used Gaussian integration with
an accuracy of $10^{-6}$.

\subsection{Pure Four-Point Function}

In Table \tref{table1} we have listed the leading hadronic
light-by-light  ${\cal O}(N_c)$ contributions
to $a_\mu$, i.e the pure four-point function in the second
column and the pseudoscalar exchange three-point-like function
in the third column,
as a function of the cut-off together with the errors quoted by VEGAS.
\begin{table}
\begin{center}
\begin{tabular}{|c|c|c|c|c|}
\hline
Cut-off &  $a_\mu$ $\times$  $10^{10}$ from &
$a_\mu$ $\times$  $10^{10}$ from
&$a_\mu$ $\times$  $10^{10}$ from
& $a_\mu$ $\times$  $10^{10}$ \\ $\mu$  &
Constituent &Pseudoscalar& $\pi^0$, $\eta$ and $\eta'$ & \\
(GeV) &Quark &Exchange ${\cal O}(N_c)$ &Exchanges& Sum \\
 &  in Figure 3(a)  & in Figure 3(b) & ${\cal O} (N_c)$ +
U(1)$_A$ & \\ \hline
0.5 & 0.78(0.01)&$-$14.2(0.1) & $-$4.8(0.1)   & $-$4.0\\
0.7 & 1.14(0.02)&$-$19.4(0.1) & $-$6.8(0.1)   & $-$5.7 \\
1.0 & 1.44(0.03)&$-$24.2(0.2) & $-$9.0(0.1)   & $-$7.6 \\
2.0 & 1.78(0.04)&$-$33.0(0.2) & $-$12.6(0.2)   & $-$10.8 \\
4.0 & 1.98(0.05)&$-$39.6(0.6) & $-$15.0(0.2) & $-$13.0\\
8.0 & 2.00(0.08)&$-$46.3(1.5) & $-$17.6(0.4)  & $-$15.6 \\
\hline
\end{tabular}
\end{center}
\caption{Results for the order $N_c$ constituent quark loop
and pseudoscalar exchange
hadronic light-by-light contributions
to $a_\mu$ in the ENJL model.}
\label{table1}
\end{table}
Since the integrand is rather irregular, this error estimate is somewhat
on the small side (see also \rcite{BB95}) and will be
largely superseded by the error in our final result.

For the bare constituent quark loop,
the result only stabilizes at a rather high value of $\mu$.
For instance, for  a bare quark loop with a constituent quark
mass of  300 MeV, the change between a cut-off of 2 GeV to a cut-off of
4 GeV is still typically 20\%.  The change from 0.7 GeV to 2 GeV
 is typically a factor of 1.8. These are the results quoted in
column 4 of Table~\tref{tablex}.
The changes for our more realistic ENJL model four-point function
can be judged from the results in
Table \tref{table1}, column 2. It still has a significant change
between 1 and 2~GeV cut-off.
This invalidates the use of any low energy
model to calculate accurately the complete
hadronic light-by-light contribution to $a_\mu$.
The bulk of these contributions does not come from the dynamics at scales
around the muon mass as it is often stated. This also explains the rather
high sensitivity to the damping provided by the vector two-point
functions as seen in \rcite{KNO85}.
Mostly due to its electric charge and heavier mass,
the contribution of the strange quark flavour
is much smaller (around $0.04\cdot 10^{-10}$)
than that of the up and down quarks shown in the Table \tref{table1}.
This value is within the quoted VEGAS error
for up and down quark contributions.
In Section \tref{7} we give
an estimate of the intermediate and high energy one constituent light
quark loop contributions and the heavier quark flavours
contributions.

\subsection{Pseudoscalar Three-Point-Like Function}
For the three-point-like  function contributions
we have done the same study of the cut-off dependence as for the
four-point function contribution. In particular we find that at large
$N_c$ the contribution of the pseudoscalar exchange
is more than one order of magnitude larger than the others.
The reason this contribution is so different can be
traced back both to the presence of two flavour anomaly vertices and
the CHPT counting.
It therefore deserves more attention. In fact, the pseudoscalar exchange
has  important next-to-leading corrections
{}from  the effects of the U(1)$_A$ anomaly that
leave the $\pi^0$ exchange as the dominant contribution to
$a_\mu^{\rm light-by-light}$.
First we give in column 3 of Table \tref{table1} the result
strictly to leading order in $1/N_c$ from the $u, d$ and $s$ flavours.

We have taken into account the effects of the U(1)$_A$
anomaly by using the physical $\pi^0$, $\eta$ and $\eta'$
mass eigenstates  as propagating states instead of
the $\bar{u}u$, $\bar{d}d$ and $\bar{s}s$ in the large $N_c$
limit.  This includes the  main effect of the
U(1)$_A$ anomaly  which is in the differences of  the masses
of the pseudoscalar $\eta$ and $\eta'$ mesons.
 In the  $\pi^0$, $\eta$ and $\eta'$
basis, the contribution from the pion intermediate
state has a charge factor $\left[\left((2/3)^2-(-1/3)^2
\right) /\sqrt{2}\right]^2$
compared to a single quark of charge one. We thus multiply the ENJL
pseudoscalar exchange result of column 3 in \rref{table1}
by this factor to obtain the $\pi^0$
contribution from the ENJL model listed in column 2 of Table \tref{tablepi0}.
\begin{table}
\begin{center}
\begin{tabular}{|c|c|c|c|c|}
\hline
Cut-off & $a_\mu\times 10^{10}$ & $a_\mu\times 10^{10}$ &
$a_\mu\times 10^{10}$ & $a_\mu\times 10^{10}$ \\
$\mu$ & & & &\\
(GeV) & ENJL &  Point-Like--VMD ($\pi^0$)& $\eta$ &$\eta'$\\
\hline
0.4 & $-$2.84(2)  &  $-$2.70(1) &$-$0.425(1) &$-$0.266(1)\\
0.5 & $-$3.70(3)  &  $-$3.46(2)&$-$0.616(2) &$-$0.399(2)\\
0.7 & $-$5.04(4)  &  $-$4.49(3)&$-$0.923(3)&$-$0.631(2)\\
1.0 & $-$6.44(7)  &  $-$5.18(3)&$-$1.180(4)&$-$0.847(3)\\
2.0 & $-$8.83(17)  &  $-$5.62(5)&$-$1.37(1) &$-$1.03(1)\\
4.0 & $-$10.51(37) &  $-$5.58(5)&$-$1.38(1)&$-$1.04(1)\\
\hline
\end{tabular}
\end{center}
\caption{\rlabel{tablepi0} The $\pi^0$ exchange
contribution to $a_{\mu}$ for the ENJL and the point-like
Wess-Zumino vertex, damped with two vector propagators for the $\pi^0$,
$\eta$ and $\eta'$.}
\end{table}

The $\eta$ and $\eta'$ contributions
we cannot estimate directly within the ENJL model. What we have done is
the following. Since the main effect is in the propagator of the
exchanged pseudoscalar meson, the ratio of the
$\pi^0$ exchange contribution to the $\eta$ or $\eta'$ contribution
has to be in good approximation model independent. So we have taken the
ratio from the point-like Wess-Zumino Lagrangian with full vector
 meson dominance and multiplied the ENJL $\pi^0$ contribution
by them to get the $\eta$ and $\eta'$ contributions.
So to get the fourth column in Table \tref{table1} we sum the last three
columns in Table \tref{tablepi0}, multiply by the 2nd column and divide
by the third one. This is our estimate for the combined
$\pi^0$, $\eta$ and $\eta'$ contributions.
 For the  calculation of the last three columns in Table \tref{tablepi0}
we have used
$P^0 \gamma \gamma$ couplings such that the experimental decay
rates $P^0 \to \gamma\gamma$ are reproduced and
a vector meson mass of 0.78~GeV. That gives ratios that
vary from 16\% at
$\mu =$ 0.4 GeV to 25\% at $\mu=$ 4.0 GeV for the ratio of the $\eta$
contribution to the $\pi^0$ one
and from 10\% at $\mu =$ 0.4 GeV to 19\% at $\mu=$ 4.0 GeV
for the ratio of the $\eta'$ contribution to the $\pi^0$ one.

We find for the pseudoscalar result
less stability at high values of the cut-off $\mu$ than for
the quark-loop contribution.
Although the change from 0.7 GeV to 2 GeV is also around 1.8, the
stability is worse for cut-off values above 4 GeV.
Notice also that the error from the integration routine VEGAS
  is larger for these values
of the cut-off. The poor stability in the pseudoscalar exchange
 is mainly due to the subtraction terms
we need to obtain the correct SU(3) flavour anomaly.
We shall give in Section \tref{7} an estimate of the
intermediate and high energy
contributions to the pseudoscalar exchange term.
We finally give in the fifth column of \rref{table1}
the sum of the second and fourth columns.

\subsection{Other Three-Point-Like Functions}
Both scalar and axial-vector exchanges in three-point-like
function contributions are much smaller than our final error.
Their results for up, down and strange quark flavours are in Table
\tref{table2}. The scalar contribution has obviously stabilized. The
axial-vector one has large cancellations and becomes numerically unstable
for a cut-off of 8~GeV. We have therefore not quoted the values for this
cut-off.
\begin{table}
\begin{center}
\begin{tabular}{|c|c|c|c|}
\hline
Cut-off &  $a_\mu$ $\times$  $10^{10}$ from &
$a_\mu$ $\times$  $10^{10}$ from
& $a_\mu$ $\times$  $10^{10}$ \\ $\mu$ &
Scalar &Axial-Vector  & \\ (GeV)
&Exchange ${\cal O}(N_c)$ &Exchange ${\cal O}(N_c)$  & Sum \\
 &  in Figure 3(b)  & in Figure 3(b) & \\ \hline
0.5 & $-$0.22(0.01)&$-$0.05(0.01) & $-$0.27 \\
0.7 & $-$0.46(0.01)&$-$0.07(0.01) & $-$0.53 \\
1.0 & $-$0.60(0.01)&$-$0.13(0.01) & $-$0.73 \\
2.0 & $-$0.68(0.01)&$-$0.24(0.02) & $-$0.92  \\
4.0 & $-$0.68(0.01)&$-$0.59(0.07)& $-$1.27 \\
\hline
\end{tabular}
\end{center}
\caption{Results for the order $N_c$
 scalar and axial-vector exchange
hadronic light-by-light contributions  to $a_\mu$ in the ENJL model.}
\label{table2}
\end{table}

\subsection{Pion and Kaon Loops
(${\cal O}(1)$ in $1/N_c$ Contributions)}
The results for the dominant contributions of order 1  in $1/N_c$
are in Table \tref{table3}.
\begin{table}
\begin{center}
\begin{tabular}{|c|c|c|c|}
\hline
Cut-off &  $a_\mu$ $\times$  $10^{10}$ from &
$a_\mu$ $\times$  $10^{10}$ from
& $a_\mu$ $\times$  $10^{10}$ \\ $\mu$ &
Pion Loop &Kaon Loop  & \\ (GeV)  &
  in Figure 3(a)  & in Figure 3(a) & Sum \\ \hline
0.5 & $-$1.20(0.03)   &$-$0.020(0.001)& $-$1.22\\
0.6 & $-$1.42(0.03)   &$-$0.026(0.001)& $-$1.45\\
0.7 & $-$1.56(0.03)   &$-$0.034(0.001)& $-$1.59\\
0.8 & $-$1.67(0.04)   &$-$0.042(0.001)& $-$1.71 \\
1.0 & $-$1.81(0.05)   &$-$0.048(0.002)& $-$1.86\\
2.0 & $-$2.16(0.06)   &$-$0.087(0.005)& $-$2.25 \\
4.0 & $-$2.18(0.07)   &$-$0.099(0.005) &$-$2.28  \\
\hline
\end{tabular}
\end{center}
\caption{Results for the order 1 in the $1/N_c$ expansion
charged pion and kaon loops hadronic light-by-light
contributions  to $a_\mu$.}
\label{table3}
\end{table}
We have saturated this contribution by
the physics of pion, kaon and rho  mesons as explained
in Section \tref{5}. Therefore, we need to verify if
these contributions really saturate for energies below the axial-vector
mass. For this, we have  studied the cut-off dependence
by varying the Euclidean cut-off $\mu$.
For these contributions we used a statistics of 10 iterations
with $10^6$  points in the Monte
Carlo routine VEGAS.
As can be seen in Table \tref{table3}, the charged
pion  loop  contribution  saturates around
2 GeV, while the kaon loop contribution saturates around 4 GeV.
{}From Table \tref{table3}, we see
that the change between the result at 1 GeV and the
result where it stabilizes is less than 20\% so we conclude
that the approximation we are doing works to this accuracy
which is good enough in view of our final uncertainty, see
Section \tref{8}. The intermediate and higher energy contributions
for this case are discussed in the next section.
The results in Table \tref{table3} are obtained using ENJL vector mesons
propagators for the vector legs.
We have also used  vector meson propagators
with a constant vector mass of 768 MeV and in this case
the charged pion plus
kaon loop contributions to $a_\mu^{\rm light-by-light}$
 saturate earlier (at $\mu=$ 0.8 GeV )
with a value around $-$1.65 $\cdot$ $10^{-10}$.

The HGS model with a=2 was used in Ref. \rcite{HKS95,HKS95b}
  to calculate this contribution. We see from \rref{HGS}
that the HGS in the non-anomalous sector and for
$a=2$ has a wrong high energy behaviour when matching QCD in the
$\pi^+ -\pi^0$ mass difference. See the negative $\nu^2$
correction to the logarithmic behaviour there.
 This also tends to lower
the contribution to $a_\mu$ too much when vector mesons are added.
This is important in the charged pion and kaon
loop contributions where the $P^+ P^- \gamma^* \gamma^*$
has an unknown high energy behaviour.
We have adopted the
criterion of using a complete VMD model inspired by the ${\cal
O}(N_c)$ ENJL model. As shown in Section \tref{5},
this does not break any Ward identity.
This choice has, at least, a good
high energy behaviour for two-point functions, e.g.
Weinberg Sum Rules are fulfilled.
This is not true for the HGS model with $a=2$.
See also \rcite{BdR91} where the $\pi^+-\pi^0$ mass difference
is calculated within this model. The result of the HGS model can however
not be excluded with these arguments.

\subsection{Sum of Low-Energy Contributions}
Adding the contributions calculated before we get the
low energy contribution to $a_\mu$ as a function of the cut-off $\mu$.
These are the final results for the low-energy contribution estimated
within the simplest version of the ENJL model. We will present other
estimates in the next section.
The results are in Table \tref{finalENJL}.
\begin{table}
\begin{center}
\begin{tabular}{|c|c|}
\hline
Cut-off & \\
$\mu$ & $a_\mu \times 10^{10}$\\
(GeV) & \\
\hline
0.5& $-$5.5(0.1)\\
0.7& $-$7.8(0.1)\\
1.0& $-$10.2(0.1)\\
2.0& $-$14.0(0.2)\\
4.0& $-$16.6(0.2)\\
\hline
\end{tabular}
\end{center}
\caption{\rlabel{finalENJL} The contribution to $a_\mu$  from
the low energy domain. The error is the one of the various contributions
given by VEGAS and added in quadrature.}
\end{table}
The ENJL model we have used is a good low energy hadronic model which
works within 20\% up to energies (0.4$\sim$0.6) GeV depending on the channel.
 We observe from the results in Table \tref{finalENJL}
 that  higher energy contributions
are certainly not negligible. The estimation of those contributions is the
subject of the next section. We conclude from this section
\be
\rlabel{pselow}
a_\mu^{\rm light-by-light}
 (\mu = 0.5 \, {\rm GeV}) = -5.5(1.5) \cdot 10^{-10}.
\ee
The error includes five times the integration error from VEGAS
and the estimated model dependence added in quadrature.
This is rather small for the
dominant pseudoscalar exchange contributions, as is shown by the small
changes in the various models presented in the next section.
At these energies
the main error is from the model dependence of the pseudoscalar
meson  ${\cal O}(1)$ contribution which we estimate to be 0.8
$\cdot$ 10$^{-10}$ to cover the results in \rcite{HKS95,HKS95b}
as mentioned above.

\section{Intermediate and High Energy Contributions}
\rlabel{7}
\setcounter{equation}{0}

In this section we estimate the hadronic contributions
{}from intermediate and high energy regions to $a_\mu^{\rm
light-by-light}$. Here we are already outside the applicability
of the ENJL model that we have only used for the low energy region.

We want to make a general comment
regarding the use of the ENJL factor
$M_V^2(-p^2)/\left(M_V^2(-p^2)-p^2\right)$ at large Euclidean  scales.
If we naively\footnote{Notice that this cannot
be done in the ENJL model since is only valid for energies
$|p| << \Lambda$.} send the Euclidean $Q^2 = - p^2 \to \infty$,
then this factor becomes 1. Therefore the photon$\times$vector propagator
in Eq. \rref{VMDsub} goes to zero as $1/Q^2$ in the ENJL model
while in a VMD model  it goes to zero as $M_V^2/Q^4$.
This difference will not affect very much
the calculation since at low energies where we apply the ENJL
model, both vector meson propagators behave very similarly.

\subsection{Pure Four-point Function }
In the case of the constituent quark loop contribution
one can still obtain an estimate of the higher
energy contributions, e.g. by
mimicking the high energy behaviour of QCD by a bare constituent
quark loop with a mass of about 1.5 GeV. This gives only an additional
correction of
$0.24(1)\cdot10^{-10}$. Using the results of \rcite{KNO85}
this scales like $1/M^2$ with $M$ the quark mass.
We have checked this behaviour
as well.
Here the mass of the heavy quark acts
as an infrared cut-off so that this heavy bare quark loop
is mimicking  the QCD behaviour
for a massless quark with an IR cut-off around 1.5 GeV.
We take this number both as the value and the uncertainty
due to the high energy region contribution for the three light flavours.
In fact we can simply assume that $\mu = M$ and add the ENJL contribution
up to the
scale $\mu$ to the bare quark-loop contribution with mass $M$ in the loop.
This leads to a total light-quark contribution of
2.2,~2.0,~1.9~and~2.0 (times $10^{-10}$) for a matching scale $\mu=M$ of
0.7,~1,~2 and 4~GeV, respectively.

We estimate the charm quark contribution with a bare quark loop.
If we damp it
with $c \bar c$ meson dominance propagators in the photon legs it
will be somewhat smaller.
This contribution is also small (about $0.2\cdot 10^{-10}$).
Therefore we obtain for the total contribution from the pure
four-point function part:
\be
\rlabel{quark}
a_\mu({\rm Quark-Loop}) = 2.1(0.3) \cdot 10^{-10}.
\ee

\subsection{Pseudoscalar Three-Point-Like Function}
Here we will only discuss
the estimate of intermediate and high energy
contributions from the pseudoscalar exchange since
this is the dominant hadronic contribution to
$a_\mu^{\rm light-by-light}$.
As seen in Section 6 all others are much smaller and
changes there will not affect our result significantly.

\begin{figure}
\begin{center}
\leavevmode\epsfxsize=10cm\epsfysize=8cm\epsfbox{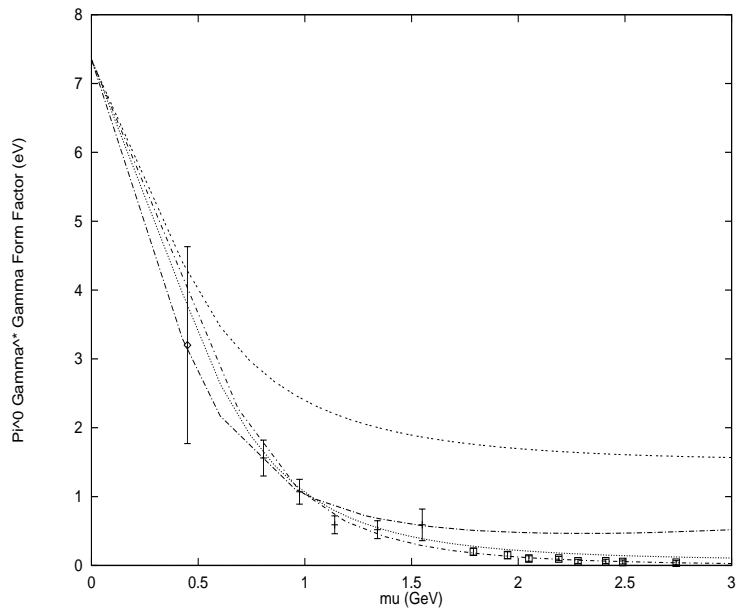}
\end{center}
\caption{The $\pi^0 \gamma^* \gamma$ form factor.
At high energies the curves from top to bottom are: ENJL, ENJL-VMD,
Point-like-VMD and Transverse-VMD. The data points are obtained from
scaling the TPC/Two-Gamma $\eta$ results \protect{\rcite{TPC2}}
(circle), the CELLO results \protect{\rcite{CELLO}}
(crosses) and CLEO-II \protect{\rcite{CLEO}} (squares).
The horizontal error bars are 0.1 GeV for the TPC/Two-Gamma
and CELLO data and 0.05 GeV for the CLEO ones.}
\label{formfactor}
\end{figure}

This contribution can be seen as the
convolution of two $P^0 \gamma^* \gamma^*$ vertices with both
$P^0$ and photons off-shell and in the Euclidean region
(see Eq. \rref{pseudoex2}).
Let us summarize what we know about this form factor.
The $\pi^0 \gamma^* \gamma$ form factor has been measured
at CELLO \rcite{CELLO} and CLEO-II \rcite{CLEO}
for values of the Euclidean invariant photon mass
$Q^2$ above (0.8)$^2$ GeV$^2$ and below (2.8)$^2$ GeV$^2$.
These are the data points in Figure \tref{formfactor}.  As we mentioned
before, we know that in QCD  the $\pi^0 \gamma^* \gamma^*$ vertex
has a $1/Q^2$ asymptotic behaviour when one of the photon legs
Euclidean  momentum is very large \rcite{GL95}. This is supported by
the phenomenological analysis of $J/ \Psi$ decays in the same
reference, where one finds that the asymptotic behaviour predicted by QCD
works reasonably well from scales around the $J/\Psi$ mass.

The chiral anomaly only fixes the $P^0 \gamma \gamma$
form factor at ${\cal O}(p^4)$, and this is fulfilled by
the ENJL model form factor.
To estimate the pseudoscalar-exchange
intermediate and high energy contribution to $a_\mu$,
what we have done is to find a phenomenological
parametrization that interpolates between the ENJL form factor,
which is supposed to work well
below 0.5 GeV,  and the measured $\pi^0 \gamma^* \gamma$
form factor for Euclidean energies above 0.5 GeV and with
its asymptotic behaviour predicted by QCD at
large Euclidean momentum. Notice that what we really need is
the $\pi^0 \gamma^* \gamma^*$ form factor as we said above.
Unfortunately no data are  available for this form factor.

Different parametrizations give quite
different contributions to $a_\mu$ for cut-offs larger than 0.5 GeV.
A lower limit would be the results from the point-like
Wess-Zumino vertex without vector meson dominance. This assumes that
the pion remains pointlike at all relevant scales.
Its
contribution to $a_\mu$  is given in column 2
of Table \tref{tablepi02} for the $\pi^0$ exchange.
\begin{table}
\begin{center}
\begin{tabular}{|c|c|c|c|c|c|}
\hline Cut-off & $a_\mu\times 10^{10}$ & $a_\mu\times 10^{10}$
&$a_\mu\times 10^{10}$&$a_\mu\times 10^{10}$&$a_\mu\times 10^{10}$ \\
$\mu$ & & &Point-Like- &Transverse- &Transverse-\\
(GeV)  &Point-like&ENJL--VMD & VMD  & VMD& VMD\\
\hline
0.5  &$-$ 4.92(2)&$-$ 3.29(2)  &$-$ 3.46(2) &$-$3.60(3)&$-$3.53(2)\\
0.7  &$-$7.68(4)&$-$ 4.24(4)  &$-$ 4.49(3) &$-$4.73(4)&$-$4.57(4)\\
1.0  &$-$ 11.15(7)&$-$4.90(5)  &$-$ 5.18(3)&$-$5.61(6)&$-$5.29(5)\\
2.0  & $-$21.3(2)&$-$ 5.63(8)  &$-$ 5.62(5)&$-$6.39(9)&$-$5.89(8)\\
4.0  &$-$32.7(5)&$-$ 6.22(17) &$-$ 5.58(5)&$-$6.59(16)&$-$6.02(10)\\
\hline
\end{tabular}
\end{center}
\caption{\rlabel{tablepi02} The $\pi^0$ exchange
contributions above $\mu =$ 0.5 GeV in
various parametrizations of the $\pi^0
\gamma^*\gamma^*$ vertex that fit the data for the $\pi^0
\gamma\gamma^*$ vertex.}
\end{table}
 The logarithmic behaviour as a function of the
cut-off is clearly visible here. The results from the ENJL model
we used for the low energy contribution are in column 2 of Table
\tref{tablepi0}. As can be seen even at a cut-off of 0.5~$GeV$ the
effect of the damping
in the ENJL model is already very important.

Both these
parametrizations (ENJL and point-like Wess-Zumino vertex)
do not fit the measured data points for
the $\pi^0\gamma^* \gamma$ form factor
above (0.5$\sim$0.6)$^2$ GeV$^2$ for the
Euclidean photon invariant mass, see Figure \tref{formfactor}.

The simple point-like
Wess-Zumino vertex plus VMD fits the $\pi^0\gamma\gamma^*$
data at high off-shellness reasonably well (see Figure
\tref{formfactor}).
This we call Point-Like--VMD parametrization and  the results
of the $\pi^0$ exchange  contribution to $a_\mu$
 are quoted in column 4 of
Table \tref{tablepi02}. This parametrization
is suppressed in both photon propagators and could be too suppressed for
both photons far off-shell. It can thus be taken as an upper
limit for the $\pi^0$ exchange of this type of contribution
for high Euclidean momenta. As another  prescription
to interpolate between the low-energy ENJL form factor and
the measured form factor we can multiply the subtraction terms
in the ENJL model $P^0 \gamma^* \gamma^*$ form factor with
  ENJL vector meson propagators,
$M_V^2(-p^2)/\left(M_V^2(-p^2)-p^2\right)$,
in all photon legs, this we call ENJL--VMD parametrization.
This is somewhat too hard at very high energies but reproduces
the form factor data at intermediate momenta reasonably well.
The $\pi^0$ exchange contribution to $a_\mu$ for this parametrization
is in column 3 of Table \tref{tablepi02}.
We observe that both Point-Like--VMD
and ENJL--VMD parametrizations give numerically
very similar contributions to $a_\mu$.
Another parametrization of the $\pi^0\gamma^*\gamma^*$ amplitude that also
fits the $\pi^0 \gamma^* \gamma$ form factor data is the following
\ba
{\cal F}^{\mu\nu}(p_1,p_2)=\frac{N_c}{6 \pi} \frac{\alpha}{f_\pi}
i \epsilon^{\mu\nu\alpha\beta} p_{1\alpha} p_{2\beta}
\tilde F(p_1^2,p_2^2,q^2)
\ea
with
\ba
\tilde F(p_1^2,p_2^2,q^2) &=&
\left( 1-g_A(-q^2) \right) \frac{M^2}{M^2-(p_1-p_2)^2} \nonumber \\
&+& \left( 1- \left( 1-g_A(-q^2) \right) \frac{M^2}{M^2-(p_1-p_2)^2}
\right) F(p_1^2,p_2^2,q^2) \nonumber \\
&\times &  \frac{M_V^2(-p_1^2)}{M_V^2(-p_1^2)-p_1^2}
\frac{M_V^2(-p_2^2)}{M_V^2(-p_2^2)-p_2^2} .
\ea
The form factor $F(p_1^2,p_2^2,q_2^2)$ is defined in Appendix
\tref{B}, Eq. \rref{anomform},
$g_A(-q^2)$ and $M_V^2(-p^2)$ in Ref. \rcite{BP94a}.
It fits the data when the ``mass'' $M$ varies
within $0.6 < M < 1.4$ GeV.The results for this parametrization
that we call Transverse-VMD  are in columns 5 and 6 of
Table \tref{tablepi02}
for a ``mass'' $M$ of $\sqrt{2}$ and 1 GeV, respectively.

In general, we obtain that the parametrizations
that fit the data also match the
ENJL low energy result at 0.5 GeV reasonably well
(compare Tables \tref{tablepi0} and \tref{tablepi02}).
This supports the low model dependence error
for the lower than $\mu=0.5$ GeV energy domain
pseudoscalar exchange contributions.

We estimated the effects of $\eta$ and $\eta'$ in the same way as
was done in the previous section.

Let us analyze  the pseudoscalar exchange
contributions from scales higher than 4 GeV.
The QCD behaviour predicted in \rcite{GL95}
goes like $1/Q^2$, where $Q^2$ is the invariant Euclidean mass of
the off-shell photon. This $Q^2$ dependence suppresses the high energy
contributions more than the point-like
Wess-Zumino vertex damped by complete VMD propagators.
In this last case we see that the contributions
{}from energies above 2 GeV are negligible. So we consider those from
above 4 GeV negligible.

We do not take the simple ENJL model as given in the previous section
but the parametrizations that fit the form factor data. Since
their results are very similar, we average
the result for the pointlike with VMD factors, the ENJL-VMD and
the transverse-VMD with $M=$ 1 GeV at 4 GeV. This is
our final result for the pseudoscalar exchange:

\be
\rlabel{pseudo}
a_\mu({\rm Pseudoscalar-Exchange})=
-8.5(1.3) \cdot 10^{-10}.
\ee
Here the error is estimated as about 15\%. This includes all the models
mentioned except the pure ENJL result.

\subsection{Other Three-Point-Like Functions}
Since the contribution of the scalar exchange
is smaller than the final error of
the dominant pseudoscalar meson exchange, we quote
for it
\be
\rlabel{scalar}
a_\mu({\rm Scalar}) = -0.68(0.2) \cdot 10^{-10}.
\ee
without making any futher estimate of the
suppressed higher energy contributions in this case.
There are in any case no pointlike subtractions here that
would have produced large possible changes.

For the axial-vector exchange contribution we have done an
analysis similar to the pseudoscalar one.
We analyzed the simplest ENJL model in Section \tref{6} and we
can do the analog of the ENJL-VMD model as well. This corresponds to removing
the VMD-factors in Eq. \rref{avvlast}.
The results for the ENJL--VMD
form factor is in Table \rref{axial2}.
\begin{table}
\begin{center}
\begin{tabular}{|c|c|}
\hline
Cut-off &  $a_\mu$ $\times$  $10^{10}$ from \\
$\mu$ &Axial-Vector \\
(GeV) &Exchange ${\cal O}(N_c)$  \\
 &  in Figure 3(b) \\ \hline
0.5 &$-$0.04(0.01) \\
0.7 &$-$0.06(0.01) \\
1.0 &$-$0.10(0.01)\\
2.0 &$-$0.15(0.01)  \\
4.0 &$-$0.35(0.04)\\
\hline
\end{tabular}
\end{center}
\caption{Results for the axial-vector exchange
in the ENJL--VMD  parametrization.}
\label{axial2}
\end{table}
Taking into account that there are cancellations which cause
VEGAS to underestimate the error at the scale of $4~GeV$, we take as
the final result for the axial exchange
\be
\rlabel{axialcon}
a_\mu({\rm Axial-Vector}) = -0.25(0.1) \cdot 10^{-10}.
\ee

\subsection{Pion and Kaon Loops
(${\cal O}(1)$ in $1/N_c$ Contributions)}
As explained in Section \tref{5}, we have saturated
the ${\cal O}(1)$ contributions  with
 charged pion and kaon loops modulated by
 vector meson propagators for the vector legs.
Since our model is
only valid for energies below the axial-vector mass, we
take the difference between the result at $\mu=$ 1 GeV and
where it stabilizes as an estimate of contributions
{}from resonances heavier than 1 GeV running in the loop.
These contributions
are higher order in the chiral counting and
 suppressed by inverse powers of the mass of these resonances.
Therefore,
we take them as an  estimate of the intermediate and high
 energy contributions for this ${\cal O}(1)$ in the $1/N_c$
 expansion contribution. For the ${\cal O}(1)$ in
the $1/N_c$ expansion contributions to $a_\mu$ we thus quote
\ba
\rlabel{eq2}
\left(a_\mu^{\rm light-by-light}\right)_{{\cal O}(1)}=
-1.9(1.3)\cdot 10^{-10}\ ,
\ea
where we have taken as central value the result at $\mu=$
1 GeV and as error the high energy contribution as estimated above
plus five times the VEGAS error
added in quadrature. There is an extra 0.8 added linearly to the
error because of model dependence, see comments in  Section \tref{6}
about its origin. It also includes an educated guess of
the ${\cal O}(1/N_c)$ corrections and
the rest of the ${\cal O}(1)$ contributions.

\section{Discussion of Results and Conclusions}
\rlabel{8}
\setcounter{equation}{0}

Our final
estimate for the hadronic light-by-light contributions
to $a_\mu$ and main result
of this work is the sum of the partial contributions
in Eqs. \rref{quark}, \rref{pseudo}, \rref{scalar},
\rref{axialcon} and \rref{eq2},
\ba
\rlabel{eq3}
a_\mu^{\rm light-by-light}=
-9.2(3.2)\cdot 10^{-10}\ .
\ea
The error is obtained by adding linearly the error of each contribution.
This is because we are essentially using the same model for all contributions
so
the error is likely to be in the same direction for all contributions.
This results in a 35\% error which we believe takes adequately
into account the model dependence error of this calculation.
This result improves and
substitutes the one in Ref. \rcite{BPP95}. There we took
as first estimate the result in Ref. \rcite{HKS95}
for the ${\cal O}(1)$ in the $1/N_c$ expansion contributions.
Here, we have given an estimate for it and
a more detailed analysis of the high energy contributions has been
performed. We also performed a more detailed study of the $\eta$ and $\eta'$
effects.

We want now to present the result in \rref{eq3} by explicitly splitting
the different contributions. First, separating the high
and the low energy contributions we get
\be
a_\mu^{\rm light-by-light} = \left(-5.0 -4.2\right) \cdot 10^{-10},
\ee
where the first number is the lower than $\mu=$ 0.5 GeV contribution
(for definiteness in the ENJL-VMD model, the others range from
$-5.0 \cdot 10^{-10}$ to $-5.5\cdot 10^{-10}$)
and the second the higher. Here one can see that the intermediate
and higher energy contributions are certainly not negligible and
$a_\mu^{\rm light-by-light}$ does not saturate at low energies
as assumed in \rcite{CNPR76,KNO85}.

It is also interesting to see how well  the $1/N_c$
expansion works in this case.
The leading $1/N_c$ result is about $-22\cdot 10^{-10}$, so the $1/N_c$
correction is about $50\%$. Notice, however, that most of it is from
the U(1)$_A$ anomaly  contribution which does not appear at order $1/N_c^2$.
The $1/N_c$ expansion works thus OK.

Finally, we see that all contributions except for the pseudoscalar
one cancel to a large extent. The part from the pseudoscalar
alone is $-8.5\cdot 10^{-10}$.
We see clearly the dominance of the pseudoscalar exchange.
Notice however the large cancellations occurring
between the four-point-like function
contribution and the pion and kaon loops ones, namely
 $(2.1-1.9)\cdot10^{-10}$. This is
possible since the pion loop is suppressed by $1/N_c$ but dominant in
the chiral counting while the other is suppressed by the chiral counting
but leading in $1/N_c$.

Since the works \rcite{CNPR76} and \rcite{KNO85}
have been updated and/or corrected by \rcite{HKS95},
we refer to this last one for the comparison with other works.
The authors of \rcite{HKS95} get as final result
\be
a_\mu^{\rm light-by-light} = -5.2(1.8) \cdot 10^{-10}.
\ee
The main differences from the results in \rcite{HKS95} are
the following. The $\eta'$, scalar and axial-vector exchange
contributions were not included there, this amounts to
\be
(-1.11 -0.68 -0.25) \cdot 10^{-10} = -2.04 \cdot 10^{-10} .
\ee
They have to be certainly included.

Another difference comes from the
different estimation of the pseudoscalar meson loop ${\cal O}(1)$
in $1/N_c$
contribution. We have essentially used complete VMD for the $P^+P^-\gamma^*
\gamma^*$ vertices for different reasons, see Sections \tref{5} and
\tref{6} for a detailed explanation.
 The numerical
difference with \rcite{HKS95} for this contribution amounts to
\be
-1.45 \cdot 10^{-10} .
\ee
We have taken into account this model dependence by adding
linearly an extra factor to the error in \rref{eq3}.

The rest of the numerical discrepancy is small
($-$0.5 $\cdot$ 10$^{-10}$ )
and  due to several causes: simplifications in \rcite{HKS95},
VEGAS numerical uncertainty, $\ldots$. Its smallness is gratifying and
reflects the low model dependence of the rest of the contributions.

Our calculation establishes that
the contribution to $a_\mu$ from light-by-light scattering
is negative and relatively large. It is one half of the
 electroweak corrections \cite{oneloop}.
This result is between two and three times the aimed experimental
 uncertainty at BNL.
Although we believe our error estimate is conservative, it has
an unsatisfactory uncertainty that will be difficult
to  reduce because of model dependence. This is mainly in
the pseudoscalar exchange and the pseudoscalar meson loop
contributions. Despite this uncertainty, the estimate in \rref{eq3}
is  still an important theoretical result
for the interpretation
of the muon $g-2$ measurement at the planned BNL experiment.

Adding the theoretical calculations of the
Standard Model  contributions to $a_\mu$ in Eqs.
\rref{oneEW}, \rref{twoEW}, \rref{QED}, \rref{EJ},
\rref{vacpolhigh} and \rref{eq3}
 gives the following
present theoretical estimate for the muon $g-2$,
\be
a^{\rm th}_\mu= 11\, 659 \, 182 (16)  \cdot 10^{-10} \, ,
\ee
where the quoted errors for different contributions are added in quadrature.
Using for the
full photon vacuum polarization insertion in the
electromagnetic muon vertex the result in \rref{AY}
instead of the one in \rref{EJ}\footnote{See Section \tref{1}
for explanation of the two different theoretical
calculations.}, gives
\be
a^{\rm th}_\mu= 11\, 659 \, 168 (11)  \cdot 10^{-10} \, .
\ee

\section*{Acknowledgments}

We thank Eduardo de Rafael for encouragement and discussions.
Useful correspondence and discussions with Profs.
M. Hayakawa, T. Kinoshita and A.I. Sanda is also acknowledged.
We thank them for drawing our attention to the form factor data.
This work was partially supported by NorFA grant 93.15.078/00.
We are grateful to
the Benasque Center for Physics where part of this work was done.
The work of EP was supported by the EU Contract Nr. ERBCHBGCT
930442.  JP thanks the
Leon Rosenfeld foundation (K\o{}benhavns Universitet) for support,
CICYT (Spain) for partial support under Grant Nr. AEN93-0234
and the DESY Theory group where part of his work was done for
hospitality.

\appendix
\def\theequation{\Alph{section}.\arabic{equation}}
\section{Construction of $\Pi^{\rho\nu\alpha\beta}$}
\rlabel{A}
\setcounter{equation}{0}

In this appendix we give the general Lorentz structure
of $\Pi^{\rho\nu\alpha\beta}(p_1,p_2,p_3)$ defined in Eq.
\rref{fourpoint}. See Fig. \tref{fig1} for definition of the
momenta. This four-point function can be decomposed by using
Lorentz covariance as follows
\ba
\fourp &\equiv& \Pi^{1}(p_1,p_2,p_3)
 g^{\rho\nu} g^{\alpha\beta} +
\Pi^{2}(p_1,p_2,p_3) g^{\rho\alpha} g^{\nu\beta}
\nonumber\\
&+&\Pi^{3} (p_1,p_2,p_3)
g^{\rho\beta} g^{\nu\alpha} \nonumber \\
&+&\Pi^{1jk}(p_1,p_2,p_3)
 g^{\rho\nu} p_j^\alpha p_k^\beta +
\Pi^{2jk}(p_1,p_2,p_3)
 g^{\rho\alpha} p_j^\nu p_k^\beta \nonumber \\
&+& \Pi^{3jk}(p_1,p_2,p_3)
 g^{\rho\beta} p_j^\nu p_k^\alpha +
\Pi^{4jk}(p_1,p_2,p_3)
 g^{\nu\alpha} p_j^\rho p_k^\beta \nonumber \\
&+& \Pi^{5jk}(p_1,p_2,p_3)
 g^{\nu\beta} p_j^\rho p_k^\alpha +
\Pi^{6jk}(p_1,p_2,p_3)
 g^{\alpha\beta} p_j^\rho p_k^\nu \nonumber \\
&+& \Pi^{ijkm}(p_1,p_2,p_3)
 p_i^\rho p_j^\nu p_k^\beta p_m^\alpha \, ,
\ea
where $i,j,k,m =$ 1, 2 or 3 and repeated indices are summed.
There are in total 138
$\Pi$-functions. Not all of them are independent since
they are related by Ward identities. In fact, when
all possible permutations of the vector legs
contributing to $\fourp$ are summed,
U(1) gauge covariance leads to the following Ward identities
\ba
\rlabel{gauge}
p_{1\nu}\fourp = p_{2\alpha}\fourp =\nonumber \\
p_{3\beta}\fourp = q{_\rho}\fourp = 0 \, .
\ea
The extensive use of these relations makes possible to
express the amplitudes $\Pi^{1,2,3}(p_1,p_2,p_3)$
and $\Pi^{1jk,\cdots,6jk}(p_1,p_2,p_3)$
in terms of $\Pi^{ijkm}(p_1,p_2,p_3)$. The fact that
we can write down everything in terms of
$\Pi^{ijkm}(p_1,p_2,p_3)$ is
again telling us that the light-by-light scattering
contribution to $a_\mu$ is a finite quantity.
We just have to calculate the UV safe
$\Pi^{ijkm}(p_1,p_2,p_3)$ amplitudes using, for
instance, a cut-off regularization scheme like proper-time.
This regulator introduces the physical cut-off
$\Lambda$ of the ENJL model, see Section \tref{3}.

The amplitudes $\Pi^{ijkm}(p_1,p_2,p_3)$
are not the minimal set
of independent amplitudes. We can still reduce it further
with some additional Ward identities. However, we
shall not use all of them and leave some ones as
checks on the resulting $\fourp$.
Since the quantity we need to compute is the antisymmetric part
of $M^{\lambda\beta}(0)$ in Eqs. \rref{MLB} and \rref{Damu},
which contains the
derivative of $\fourp$ with respect to $p_{3\lambda}$ at $p_3=0$,
we can reduce the number of needed amplitudes to $\Pi^{3jkm}
(p_1,p_2,p_3)$,
$\Pi^{i3km}(p_1,p_2,p_3)$,  $\Pi^{ij3m}(p_1,p_2,p_3)$
and the derivatives of
$\Pi^{ijk1}(p_1,p_2,p_3)-\Pi^{ijk2}(p_1,p_2,p_3)$
with respect to $p_{3\lambda}$
at $p_3=0$. Here $i,j,k,m=$ 1 or 2, so that we need 32 functions.
This is the set of amplitudes which we will use in all
our calculations.

\section{Three-Point Functions}
\rlabel{B}
\setcounter{equation}{0}

In this appendix we give the three-point functions
needed in Section \tref{4}.
Barred three-point functions are the one-constituent-quark loop ones.
{}From Eqs. \rref{threescalar}, \rref{pseudoex2} and \rref{threeaxial}
we see that we only
need barred three-point functions, therefore we only give
the explicit expression for them. The full three-point functions
can be obtained using the methods explained in \rcite{BP94a}
in a straightforward manner. We only will give the
contribution to the three-point function given by the clock-wise
orientation of the internal quark lines. The other orientation
is taken into account in the permutation of the external vector
legs, see Section \tref{2}.

Let us start with the $SVV$ three-point function. This is defined as
\ba
\Pi^{SVV}_{\mu\nu}(p_1,p_2) &\equiv&
i^2 {\dis \int} {\rm d}^4 x  {\dis \int} {\rm d}^4 y
e^{i(p_1 \cdot x +p_2 \cdot y)} \,
\langle 0| T \left( S^{ij}(0) V^{kl}_\mu(x) V^{mn}_\nu(y)
\right) |0 \rangle ,\nonumber \\
\ea
with $V^{ij}_\mu(x)\equiv [\bar q_i (x) \gamma_\mu q_j (x)]$ and
$S^{kl}(y)\equiv - [\bar q_k(x) q(x)_l]$. Summation over colour
between brackets is understood and latin indices are flavour
indices.
Owing to Lorentz covariance
this three-point function can be decomposed as follows
\ba
\Pi^{SVV}_{\mu\nu}(p_1,p_2) &\equiv&
\Pi_1(p_1,p_2) \, p_{1\mu} p_{1\nu} +
\Pi_2(p_1,p_2) \, p_{1\mu} p_{2\nu} \nonumber \\
&+& \Pi_3(p_1,p_2) \, p_{2\mu} p_{1\nu} +
\Pi_4(p_1,p_2) \, p_{2\mu} p_{2\nu} +
\Pi_5(p_1,p_2) \, g_{\mu\nu} \,  . \nonumber \\
\ea
We shall use the Ward identities for three-point functions to
write down $\Pi^5(p_1,p_2)$ in terms of the
$\Pi^i(p_1,p_2)$, $i=1,\cdots,4$ amplitudes
which are UV finite. We compute the corresponding barred functions
$\overline{\Pi}^i(p_1,p_2)$, $i=1,\cdots,4$
with the standard Feynman parametrization
technique and using proper-time regularization.
This regulator introduces the physical cut-off
$\Lambda$, see Section \tref{3}.
The $VVS$ three-point function can be obtained from this one
using the identity
\be
\Pi^{VVS}_{\mu\nu}(p_1,p_2)=\Pi^{SVV}_{\mu\nu}
(-(p_1+p_2),p_1) \, .
\ee

The anomalous $PVV$ three-point function is defined as
\ba
\Pi^{PVV}_{\mu\nu}(p_1,p_2) &\equiv&
i^2 {\dis \int} {\rm d}^4 x  {\dis \int} {\rm d}^4 y
e^{i(p_1 \cdot x +p_2 \cdot y)} \,
\langle 0| T \left( P^{ij}(0) V^{kl}_\mu(x) V^{mn}_\nu(y)
\right) |0 \rangle \nonumber \\
\rlabel{BARPVV}
\ea
with $P^{ij}(x) \equiv [ \bar q_i (x) i \gamma_5 q_j (x)]$.
It was calculated in the ENJL model  in Ref.
\rcite{BP94a}. With the same notation as there, we get
\ba
{\overline \Pi}^{+}_{\mu\nu}(p_1,p_2) &=&
\frac{N_c}{16 \pi^2} \, \epsilon_{\mu\nu\beta\rho}
p_1^\beta p_2^\rho \, F(p_1^2,p_2^2,q^2) \, {2 \over M_i} \, ,
\rlabel{BARME}
\ea
with $M_i$ is the flavour $i$ constituent quark mass.
Function \rref{BARME} is the one-loop constituent
quark (barred function) contribution with clock-wise orientation of
the internal quark line to the three-point function \rref{BARPVV}.
The function $F(p_1^2,p_2^2,q^2)$ is
\ba
\rlabel{anomform}
F(p_1^2,p_2^2,q^2)&=& 1  - \hat I_3(M_Q^2,0,0,0)  +
\hat I_3(M_i^2,p_1^2,p_2^2,q^2) , \,  ,
\nonumber \\
\ea
where $M_Q$ is the constituent quark mass  in the
chiral limit and
\ba \rlabel{i3}
\hat I_3(M_i^2,p_1^2,p_2^2,q^2) &\equiv&
2 M_i^2 {\dis \int^1_0}
{\rm d} x  {\dis \int^{1-x}_0} {\rm d}y
\frac{\Gamma_1(M^2(x,y)/\Lambda^2)}{M^2(x,y)}
\ea
where
\be
\rlabel{m2}
M^2(x,y) \equiv M_i^2 - x(1-x) p_1^2 - y(1-y) p_2^2
- 2 xy p_1 \cdot p_2 \, ,
\ee
 $\Lambda$ is the physical cut-off, see Section \tref{3}
and
\be
\Gamma_1(\epsilon) = e^{-\epsilon} \, .
\ee
Notice that in Eq. \rref{anomform},  we have given
the constituent quark mass dependence that was not
explicit in Ref. \rcite{BP94a}. In Eq. \rref{i3} we have
corrected an  obvious misprint in Ref. \rcite{BP94a}.

Again the $VVP$ three-point function can be obtained
{}from the $PVV$ one by symmetry
\be
\Pi^{VVP}_{\mu\nu}(p_1,p_2)=\Pi^{PVV}_{\mu\nu}
(-(p_1+p_2),p_1) \, .
\ee

The anomalous $AVV$ three-point function is defined as
\ba
\Pi^{AVV}_{\mu\nu\alpha}(p_1,p_2) &\equiv&
i^2 {\dis \int} {\rm d}^4 x  {\dis \int} {\rm d}^4 y
e^{i(p_1 \cdot x +p_2 \cdot y)} \,
\langle 0| T \left( A^{ij}_\mu(0) V^{kl}_\nu(x) V^{mn}_\alpha(y)
\right) |0 \rangle ,\nonumber \\
\ea
with $A^{ij}_\mu(x) \equiv [ \bar q_i (x) \gamma_\mu
 \gamma_5 q_j (x) ]$.
The most general Lorentz decomposition
in four dimensions of this three-point function is
\ba
\Pi_{AVV}^{\mu\nu\alpha}(p_1,p_2)&\equiv&
 i\, \epsilon^{\mu\nu\alpha\beta}
(p_{1\beta}\Pi_1(p_1,p_2)+p_{2\beta}\Pi_2(p_1,p_2)) \nonumber
\\ &+&
 i\, \epsilon^{\mu\nu\gamma\delta}p_{1\gamma}p_{2\delta}
\left( p_1^\alpha\Pi_3(p_1,p_2)+p_2^\alpha \Pi_4(p_1,p_2) \right)
\nonumber \\ &+&
 i\, \epsilon^{\mu\alpha\gamma\delta}p_{1\gamma}p_{2\delta}
\left( p_1^\nu\Pi_5(p_1,p_2)+p_2^\nu\Pi_6(p_1,p_2) \right)   \, .
\ea
We have used Schouten identities to eliminate redundant terms.
These identities can be also used to relate this basis to the one
used in \rcite{AD69}. As for the other three-point functions
not all the amplitudes $\Pi_i(p_1,p_2)$, $i=1, \cdots, 6$,
are independent
since they are related through Ward identities.
In this case there are three Ward identities that reduce
the six amplitudes to three independent ones.
For the diagonal flavour case, the Ward identities for the
two vector legs,  give
\ba
\Pi_1(p_1,p_2) &=&p_2^2\Pi_4(p_1,p_2) +
p_1 \cdot p_2\Pi_3(p_1,p_2)
  \nonumber\\
\Pi_2(p_1,p_2) &=&p_1^2\Pi_5(p_1,p_2) +p_1
\cdot p_2 \Pi_6(p_1,p_2) .
\ea
The Ward identity on the axial leg relates the amplitudes
$\Pi_3(p_1,p_2)$, $\Pi_4(p_1,p_2)$, $\Pi_5(p_1,p_2)$
and $\Pi_6(p_1,p_2)$ to the form factor of the
$\Pi^{PVV}_{\mu\nu}(p_1,p_2)$ anomalous three-point function
in \rref{anomform}.
The explicit expressions for the corresponding
barred $AVV$ three-point function amplitudes (pulling out
the flavour structure factor $\delta^{lm} \delta^{ni}
\delta^{jk}$) are
\ba
\overline \Pi_3(p_1,p_2)&=& - \overline \Pi_6(p_1,p_2) \nonumber \\
&=&\frac{N_c}{16\pi^2} \, 8 \, {\dis \int^1_0}
{\rm d}x {\dis \int^{1-x}_0} {\rm d}y \,  {xy \over M^2(x,y)}
\, {\tilde \Gamma}_1(M^2(x,y)/\Lambda^2) \, ;\nonumber \\
\overline \Pi_4(p_1,p_2)&=&
\frac{N_c}{16\pi^2} \, 8 \, {\dis \int^1_0}
{\rm d}x {\dis \int^{1-x}_0} {\rm d}y \,  {y(1-y) \over M^2(x,y)}
\, {\tilde \Gamma}_1(M^2(x,y)/\Lambda^2) \, ; \nonumber \\
\overline \Pi_5 (p_1,p_2) &=&
-\frac{N_c}{16\pi^2} \, 8 \, {\dis \int^1_0}
{\rm d}x {\dis \int^{1-x}_0} {\rm d}y \,  {x(1-x) \over M^2(x,y)}
\, {\tilde \Gamma}_1(M^2(x,y)/\Lambda^2) \,  , \nonumber \\
\ea
with $M^2(x,y)$ defined in Eq. \rref{m2} and
\ba
\rlabel{avvlast}
{\tilde \Gamma}_1 (M^2(x,y)/\Lambda^2) &=&
\frac{M^2(x,y)}{M^2(x,y)-M_i^2} \nonumber \\
&\times& \left[ \Gamma_1 (M^2_Q/\Lambda^2) -1
+ \frac{(M_V^2(-p_1^2)-p_1^2)(M_V^2(-p_2^2)-p_2^2)}
{M_V^2(-p_1^2)M_V^2(-p_2^2)} \right] \nonumber \\
&-& \frac{M_i^2}{M^2(x,y)-M_i^2} \Gamma_1 (M^2(x,y)/\Lambda^2)
\, . \nonumber \\
\ea
The function $M_V^2(-p^2)$ can be found in \rcite{BP94a}
and each $M_V^2(-p_i^2)$ in \rref{avvlast}
corresponds to the vector leg flavour numbers in \rref{BARPVV}.
Again the $VVA$ three-point function can be obtained
{}from the $AVV$ one by symmetry
\be
\Pi^{VVA}_{\mu\nu\alpha}(p_1,p_2)=\Pi^{AVV}_{\alpha\mu\nu}
(-(p_1+p_2),p_1) \, .
\ee

\end{document}